\documentclass[11pt]{article}

\usepackage{graphicx}
\usepackage{amsmath}
\usepackage{amsfonts}
\usepackage{amssymb}
\usepackage{graphicx}
\usepackage{wrapfig}
\usepackage{rotating}

\def\be{\begin{eqnarray}}
\def\ee{\end{eqnarray}}
\def\nn{\nonumber}

\def\tr{{\rm tr}\,}

\hoffset= - 0.7in         

\title{{\bf A direct proof of AGT conjecture at $\beta = 1$ }
\vspace{.2cm}}
\author{{\bf A.Mironov}\footnote{ {\small {\it
Lebedev Physics Institute} and {\it ITEP, Moscow, Russia}};
mironov@itep.ru; mironov@lpi.ru}, {\bf A.Morozov}\thanks{{\small
{\it ITEP, Moscow, Russia} and {\it
Laboratoire de Mathematiques et
Physique Theorique, CNRS-UMR 6083, Universite Francois Rabelais de
Tours, France}}; morozov@itep.ru} \ and {\bf
Sh.Shakirov}\thanks{{\small {\it ITEP, Moscow, Russia} and {\it
MIPT, Dolgoprudny, Russia}}; shakirov@itep.ru}\date{ }}

\begin{document}

\maketitle

\vspace{-6.0cm}

\begin{center}
\hfill FIAN/TD-17/10\\
\hfill ITEP/TH-59/10\\
\end{center}

\vspace{3.5cm}

\begin{abstract}
The AGT conjecture claims an equivalence of conformal blocks in 2d CFT
and sums of Nekrasov functions (instantonic sums in 4d SUSY gauge theory).
The conformal
blocks can be presented as Dotsenko-Fateev $\beta$-ensembles, hence, the AGT
conjecture implies the equality between Dotsenko-Fateev $\beta$-ensembles
and the Nekrasov functions. In this paper, we prove it in a particular case of $\beta=1$
(which corresponds to $c = 1$ at the conformal side and to
$\epsilon_1 + \epsilon_2 = 0$ at the gauge theory side) in a very direct way.
The central role is played by representation
of the Nekrasov functions through correlators of characters (Schur polynomials)
in the Selberg matrix models. We mostly concentrate on the case of $SU(2)$
with 4 fundamentals, the extension to other cases being straightforward.
The most obscure part is extending to an
arbitrary $\beta$: for $\beta \neq 1$, the Selberg integrals that we use do not
reproduce single Nekrasov functions, but only sums of them.
\end{abstract}


\section{Introduction}

One of the most recent instructive discoveries in string theory,
the \emph{AGT conjecture} \cite{AGT} (see \cite{MMu3}-\cite{AlbaLit} for later progress)
states an equivalence between conformal blocks
in two-dimensional conformal field theory (with $W_N$ symmetry) on one
side \cite{CFT}, and the LMNS instanton partition functions \cite{LMNS} in
four-dimensional supersymmetric ($SU(N)$) gauge theory on the other.
This relation is important for
several reasons. Basically, it provides a very explicit (and rigorously formulated)
realization of the string theory idea for a similarity between 4d supersymmetric and
2d conformal field theories, much more concrete than the standard AdS/CFT duality.
Serving as a bridge between two different fields of research, the AGT relation
stimulates progress in the both of them (say, activates the once abandoned studies of
conformal blocks of $W_N$-algebras). It also provides \cite{surop} an advanced version
\cite{NS} of the
well-known correspondence \cite{SWint} between 4d effective low-energy actions
(Seiberg-Witten prepotentials \cite{SW}) and integrable systems (often formulated in
terms of 2d bosons and fermions).

Remarkably, apart from the two initial branches of physics connected by the AGT
relation, there is still another, third field of research, which gets naturally
involved: the theory of matrix models \cite{DV,Wyl,AGTmamo,ito,MMMS}.
This was, of course, expected from the very beginning that matrix models belong to the
same level of complexity as Seiberg-Witten prepotentials, their partition functions
are long known to provide solutions to classical integrable hierarchies
\cite{mamoint}, etc
(see \cite{SWMM} for an exact correspondence between matrix models and Seiberg-Witten
theory). Nowadays these expectations turned
into a very clearly formulated statement: that matrix models provide explicit
integral representations for the conformal blocks.
To be more precise, these are integral representations of conformal blocks $B(q)$
in the non-trivially interacting 2d CFT in terms of correlators with screening charge
insertions in the free field 2d CFT \emph{a-la}
Dotsenko and Fateev \cite{DF}. A representative example is the four-point spherical
conformal block (related to $SU(2)$ Nekrasov
function with four fundamental matter hypermultiplets) \cite{MMMS}:
\begin{align*}
\mbox{4-point conformal block} = {\cal B}\big( q \big) = \mbox{integrated free-field correlator} =
\end{align*}
\begin{align*}
= \left<\left< :e^{\tilde\alpha_1\phi(0)}:\ :e^{\tilde\alpha_2\phi(q)}:\
:e^{\tilde\alpha_3\phi(1)}:\ :e^{\tilde\alpha_4\phi(\infty)}:\
\left(\int_0^q :e^{b\phi(z)}:\,dz\right)^{N_1}
\left(\int_0^1 :e^{b\phi(z)}:\,dz\right)^{N_2}\right>\right> =
\end{align*}
\begin{align}
= q^{{\alpha_1 \alpha_2}/{2 \beta}} \
(1 - q)^{{\alpha_2 \alpha_3}/{2 \beta}}
\prod\limits_{i = 1}^{N_1} \int\limits_{0}^{q} d z_i \
\prod\limits_{i = N_1+1}^{N_1 + N_2} \int\limits_{0}^{1} d z_i \
\prod\limits_{i < j} (z_j - z_i)^{2 \beta}
\prod\limits_{i} z_i^{\alpha_1} (z_i - q)^{\alpha_2} (z_i - 1)^{\alpha_3}
\label{Fateev-Dotsenko}
\end{align}
\smallskip\\
where $\tilde\alpha_i=\alpha_i/2b$ and $\beta=b^2$. The second line is
the correlator of normally ordered chiral
vertex operators, corresponding to the initial four external fields and additional
$N_1+N_2$ screening charges, inserted in positions $z_1, \ldots, z_{N_1+N_2}$ and
integrated with peculiar choices of integration contours.
Such correlators are free field (Gaussian) averages, straightforwardly evaluated with help of the Wick theorem

\begin{align}
\left<\left< :e^{\tilde\alpha_1\phi(z_1)}: \ \ldots \
:e^{\tilde\alpha_m\phi(z_m)}: \right>\right> =
\prod\limits_{ 1 \leq i < j \leq m} (z_j - z_i)^{2\tilde\alpha_i \tilde\alpha_j}
\end{align}
\smallskip\\
and finally put into the form of multiple integral (\ref{Fateev-Dotsenko}) similar to
matrix model eigenvalue integrals. For a generic $\beta \neq 1$, determined
by the value of screening charge $b$ and related to the central charge via
$c = 1 - 6(b-1/b)^2$, the integral is not, strictly speaking, an ordinary matrix
model, it is rather a generalization known as $\beta$-ensemble \cite{beta,MMStalk} or
"conformal" matrix model \cite{confmamo,ito}. The difference, however, is not too
drastic: it is well-known that matrix model theory is easily generalizable
from $\beta = 1$ to arbitrary values of $\beta$, see \cite{MMStalk} for a recent
summary.

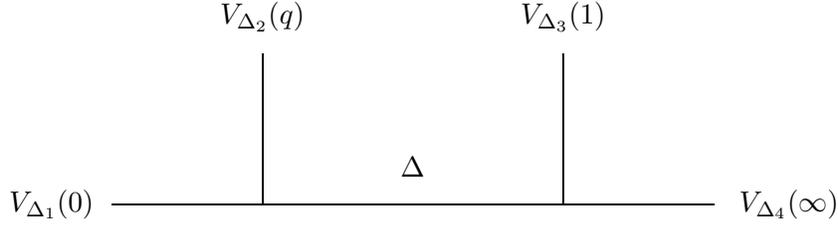
\begin{figure}\begin{center}
\unitlength 1mm 
\linethickness{0.4pt}
\ifx\plotpoint\undefined\newsavebox{\plotpoint}\fi 
\begin{picture}(100,20)(0,8)
%
%
\put(10,10){\line(1,0){80}}
\put(30,10){\line(0,1){20}}
\put(70,10){\line(0,1){20}}
\put(2,10){\makebox(0,0)[cc]{$V_{\Delta_1}(0)$}}
\put(30,35){\makebox(0,0)[cc]{$V_{\Delta_2}(q)$}}
\put(70,35){\makebox(0,0)[cc]{$V_{\Delta_3}(1)$}}
\put(100,10){\makebox(0,0)[cc]{$V_{\Delta_4}(\infty)$}}
\put(50,15){\makebox(0,0)[cc]{$\Delta$}}
\end{picture}
\caption{\footnotesize{
Feynman-like diagram for the 4-point conformal block. The external legs represent
primary fields in 2d CFT; the structure of the graph shows the order of
contractions in the operator product expansion procedure.
}}\label{4block}
\end{center}\end{figure}

There are many different conformal blocks classified by the three main
characteristics: a) conformal diagram, i.e. a graph with external legs,
which shows the order of their OPE contraction; b) genus of underlying
Riemann surface and c) rank $N$ of the symmetry, which is $N=2$ for the usual
Virasoro conformal blocks, and higher $N$ for conformal blocks of $W_N$ algebras.
For all of them, the Dotsenko-Fateev integrals can be straightforwardly written:
extra internal dimensions are described by adding screening operators with different
integration contours \cite{Wyl,MMMS}; higher genera surfaces are described by substitution
of free field Green functions by appropriate theta-functions \cite{DV,DFhg}; higher rank
symmetries are described by making $\alpha$'s and $b$'s vector-valued
\cite{confmamo}.
Because of this, and also because of their natural simplicity, it is convenient
to use the Dotsenko-Fateev integrals to represent the whole variety of conformal blocks
in the left hand (conformal) side of the AGT conjecture. This is exactly what we do
in the present paper: we use for the conformal blocks the matrix model
Dotsenko-Fateev representation \cite{DV,Wyl,AGTmamo,ito,MMMS}.

On the other (gauge theory) side of the AGT conjecture, there are Nekrasov
functions, the ultimate outcome of evaluation of integrals
over the instanton moduli spaces in ${\cal N} = 2$ SUSY Yang-Mills theories \cite{NF}.
Since integrals over instanton moduli spaces typically diverge, they need to be
regularized, and this, as usual, can be done in many different ways. One of the most
popular ways to regularize these integrals \cite{LMNS} relies
on introduction of the so-called $\Omega$-background and associated deformation
parameters $\epsilon_1,\epsilon_2$. The integrals over moduli spaces, regularized
in this way, were evaluated in \cite{NF} and
finally represented as series in instanton parameters, with all terms explicit.

\begin{wrapfigure}{l}{230pt}
  \begin{center}
\vspace{-5ex} \includegraphics[width=210pt]{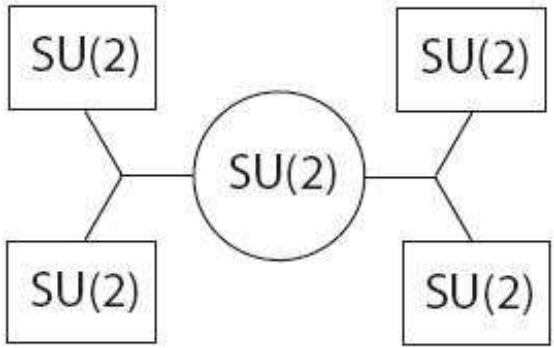}
  \end{center}
  \caption{Quiver diagram for the $SU(2)$ Nekrasov functions with four fundamentals.
  The external boxes represent the matter hypermultiplets, the central circle represents
  the gauge group, and the structure of the graph shows transformation properties
  of the matter hypermultiplets under the gauge group action.}
\end{wrapfigure}

There are many different types of Nekrasov functions, classified
according to quiver diagrams \cite{NFquiver,AGT}, i.e. graphical
representations of the field content of a given theory with detailed
indication of gauge groups and transformation properties of the
matter multiplets. According to the AGT conjecture \cite{AGT}, each
of these types of Nekrasov functions corresponds to a conformal
block: the conformal diagram can be simply read off from the quiver
diagram, with genus corresponding to the number of loops and with
the symmetry (Virasoro or, generally, $W_N$) fixed by rank of the
gauge group ($SU(2)$ or, generally, $SU(N)$). Such a "dictionary"
between 2d and 4d theories extends the one, originally suggested in
\cite{SWint,SWd,SWf}, and represents one of the most explicit manifestations of the
gauge-string duality over the last decades.

It is natural that, apart from generalizations and possible applications, more
and more attention is getting attracted to the questions of understanding and
proof of the AGT conjecture. The \emph{understanding} of the otherwise mysterious
connection between 2d and 4d theories is generally believed to be based upon
existence of a certain unique 6d theory, which is in charge (through
compactification) of the AGT relation. However, due to technical complications
this direction remains largely philosophical, and has been unable to produce a
\emph{proof} yet.

Since the AGT relation is essentially the equality between
the Nekrasov functions and Dotsenko-Fateev integrals,
a more concrete approach could be to make use of the well-developed methods of matrix models for the proof. Several suggestions have been proposed on
how to deal with the Nekrasov functions within the matrix model framework
\cite{DV,Wyl,AGTmamo,ito,MMMS}.

In \cite{MMStalk} in order to proof the AGT conjecture in a more concrete way,
we suggested to use that the Nekrasov functions are
$\epsilon_1,\epsilon_2$-deformations of the celebrated Seiberg-Witten
prepotentials, and the corresponding Seiberg-Witten theory
coincides with the Seiberg-Witten theory of the planar limit of the
Dotsenko-Fateev matrix model \cite{DV,Wyl,AGTmamo}. Then, one may restore the
$\epsilon_1,\epsilon_2$-deformations of the both Seiberg-Witten theories
by the topological recursion \cite{AMMEO}, so that they still would coincide,
with the Seiberg-Witten differential in the recursion being given by
the exact 1-point resolvent of the matrix model (or, more precisely, of the
$\beta$-ensemble). Another possibility is to use the Harer-Zagier recursion
\cite{HZ,MMStalk}. However, at the moment too little is known about
matrix model representation of the Nekrasov functions, thus, this program remains
to be accomplished. Development of the Harer-Zagier technique may play an important role here.

In this paper, we suggest to look at Nekrasov functions literally: as
explicitly known sums over partitions (Young diagrams, see Figure 3).Such series are indeed available for various \begin{wrapfigure}{r}{210pt}
  \begin{center}
\vspace{-4ex} \includegraphics[width=190pt]{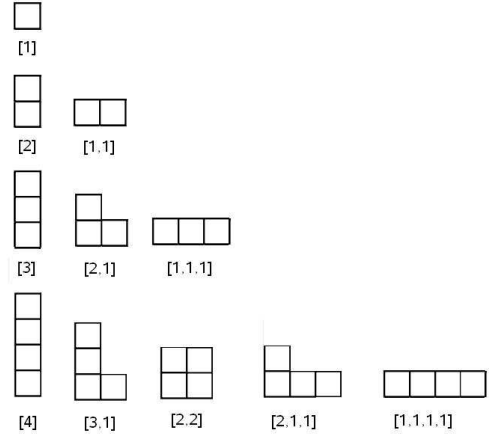}
  \end{center}
\vspace{-4ex}
  \caption{Several first Young diagrams.}
\end{wrapfigure}
Nekrasov functions in the literature \cite{NF,NFmore}.
For example, in the case of $SU(2)$ with four
fundamental hypermultiplets (related to the 4-point spherical conformal block
(\ref{Fateev-Dotsenko})) the Nekrasov function can be written as a sum
\begin{align}
Z_{Nek}(q) \ = \ & \sum\limits_{A,B} \ N_{A,B} \ q^{|A| + |B|}
\label{NekFull}
\end{align}
over the Young diagrams $A = [A_1 \geq A_2 \geq \ldots]$ and
$B = [B_1 \geq B_2 \geq \ldots]$, with the coefficients $N_{A,B}$ being rational
functions of the masses $\mu_1, \mu_2, \mu_3, \mu_4$, of the Coulomb parameter
$a$ and of the deformation parameters $\epsilon_{1,2}$. Explicitly, several first coefficients
$N_{AB}$ have the form
\vspace{-1ex}
\begin{align}
N_{[1][]} = -\frac{1}{\epsilon_1\epsilon_2}\cdot
\frac{\prod_{r=1}^4 (a + \mu_r)}
{2a(2a+\epsilon_1+\epsilon_2)}, \ \ N_{[][1]} = -\frac{1}{\epsilon_1\epsilon_2}\cdot
\frac{\prod_{r=1}^4 (a - \mu_r)}
{2a(2a-\epsilon_1-\epsilon_2)}
\label{NekMasses1}
\end{align}
\vspace{-3ex}
\begin{align}
N_{[1][1]} = \frac{1}{\epsilon_1^2\epsilon_2^2}\cdot
\frac{\prod_{r=1}^4 (a + \mu_r)(a-\mu_r)}
{(4a^2-\epsilon_1^2)(4a^2-\epsilon_2^2)}
\label{NekMasses2}
\end{align}
\vspace{-3ex}
\be
N_{[2][]} = \frac{1}{2!\,\epsilon_1\epsilon_2^2
(\epsilon_1-\epsilon_2)}\cdot
\frac{\prod_{r=1}^4 (a + \mu_r)(a+\mu_r+\epsilon_2)}
{2a(2a+\epsilon_2)(2a+\epsilon_1+\epsilon_2)(2a+\epsilon_1+2\epsilon_2)},
\label{NekMasses3}
\\
N_{[][2]} = \frac{1}{2!\,\epsilon_1\epsilon_2^2
(\epsilon_1-\epsilon_2)}\cdot
\frac{\prod_{r=1}^4 (a - \mu_r)(a-\mu_r-\epsilon_2)}
{2a(2a-\epsilon_2)(2a-\epsilon_1-\epsilon_2)(2a-\epsilon_1-2\epsilon_2)},
\label{NekMasses4}
\\
N_{[11][]} = -\frac{1}{2!\,\epsilon_1^2\epsilon_2
(\epsilon_1-\epsilon_2)}\cdot
\frac{\prod_{r=1}^4 (a + \mu_r)(a+\mu_r+\epsilon_1)}
{2a(2a+\epsilon_1)(2a+\epsilon_1+\epsilon_2)(2a+2\epsilon_1+\epsilon_2)},
\label{NekMasses5}
\\
N_{[][11]} = -\frac{1}{2!\,\epsilon_1^2\epsilon_2
(\epsilon_1-\epsilon_2)}\cdot
\frac{\prod_{r=1}^4 (a - \mu_r)(a-\mu_r-\epsilon_1)}
{2a(2a-\epsilon_1)(2a-\epsilon_1-\epsilon_2)(2a-2\epsilon_1-\epsilon_2)}
\label{NekMasses6}
\ee
and so on (omitting the trivial $N_{[][]}=1$). For an explicit formula for the generic $N_{AB}$, see (\ref{NekCoefMasses}).

This "sum-over-partitions" point of view allows one to make a direct
contact with matrix models, where these sums appear after
\emph{character expansion} \cite{charex}: decomposition of the
integrand in a proper basis of symmetric polynomials. At $\beta =
1$, the proper basis is realized by the ordinary Schur polynomials,
i.e. by the $GL(\infty)$ characters, which are labeled by
partitions:
\begin{align}
\chi_{1}(p) = p_1
\end{align}
\vspace{-2ex}
\begin{align}
\chi_{2}(p) = \dfrac{p_1^2}{2}+\dfrac{p_2}{2}, \ \ \
\chi_{11}(p) = \dfrac{p_1^2}{2}-\dfrac{p_2}{2}
\end{align}
\vspace{-2ex}
\begin{align}
\chi_3(p) = \dfrac{p_3}{3} + \dfrac{p_1 p_2}{2} + \dfrac{p_1^3}{6},
\ \ \
\chi_{21}(p) = -\dfrac{p_3}{3}  + \frac{p_1^3}{3}, \ \ \
\chi_{111}(p) = \dfrac{p_3}{3} - \dfrac{p_1p_2}{2} + \frac{p_1^3}{6}
\end{align}
\smallskip\\
etc., where $p_k = \sum z_i^k$ are the power sums. For $\beta \neq 1$, the proper
deformation of the Schur polynomials is the Jack polynomials (aka $\beta$-characters),
which depend on $p_k$ and on a single additional parameter $\beta$.
Further deformations, to the McDonald and, generally, Askey-Wilson polynomials,
depend on more additional parameters and are relevant for description of 5d
 \cite{Awata} and,
perhaps, 6d gauge theories.

To interpret series (\ref{NekFull}) as a character expansion, one needs to express
the Nekrasov coefficients $N_{A,B}$ through the Schur polynomials. In this paper,
we describe a solution to this problem for $\beta = 1$, which corresponds to the
case of $\epsilon_1 + \epsilon_2 = 0$ for the Nekrasov function (the minus in the argument
of the Schur function corresponds to the transposed Young diagram, see (\ref{Inversion})):

\begin{figure}
  \begin{center}
\includegraphics[width=250pt]{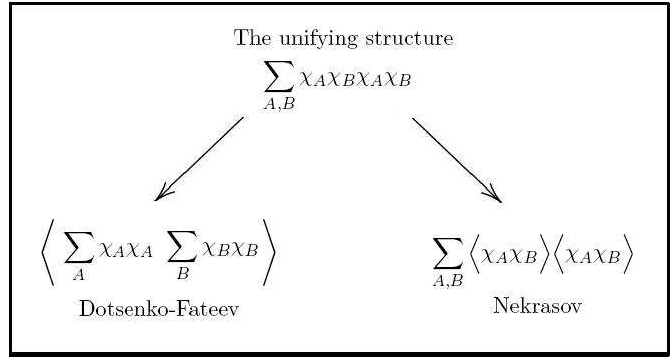}
  \end{center}
  \caption{The core idea of the proof. This is a typical duality, when one unifying structure
  $(( \chi_A\chi_B\chi_A\chi_B ))$
decomposes into two different channels $(\chi_A\chi_A)(\chi_B\chi_B)$ and
$(\chi_A\chi_B)(\chi_A\chi_B)$.}
\end{figure}
\begin{equation}
\addtolength{\fboxsep}{5pt}
\boxed{
\begin{gathered}
N_{A,B} \Big( \epsilon_1 = \hbar, \epsilon_2 = - \hbar \Big) = \Big< \chi_{A}\big( - p_k - v_+ \big) \ \chi_B\big(p_k\big) \Big>_+ \ \ \Big< \chi_A\big( p_k \big) \ \chi_{B}\big( -  p_k - v_- \big) \Big>_-
\end{gathered}
}\label{CharacterNekrasov}
\end{equation}
\smallskip\\
where the brackets $\big< \big>_+,\big< \big>_-$ denote averaging over two Selberg ensembles
\begin{align}
\Big< f \Big>_\pm \ = \ \dfrac{
\int\limits_{0}^{1} dz_1 \ldots \int\limits_{0}^{1} dz_{N_\pm} \prod\limits_{i<j} (z_i - z_j)^{2} \prod\limits_{i} z_i^{u_\pm} (z_i - 1)^{v_\pm} \ f\big(z_1, \ldots, z_{N_\pm}\big)}{\int\limits_{0}^{1} dz_1 \ldots \int\limits_{0}^{1} dz_{N_\pm} \prod\limits_{i<j} (z_i - z_j)^{2} \prod\limits_{i} z_i^{u_\pm} (z_i - 1)^{v_\pm} }
\label{SelbergAverage}
\end{align}
with

\begin{align}
u_+ = \dfrac{\mu_2 - \mu_1}{\hbar}, \ \ u_- = \dfrac{\mu_4 - \mu_3}{\hbar}
\label{SelbergParameters1}
\end{align}
\begin{align}
v_+ = \dfrac{\mu_1 + \mu_2}{\hbar}, \ \ v_- = \dfrac{\mu_3 + \mu_4}{\hbar}
\label{SelbergParameters2}
\end{align}
\begin{align}
N_+ = \dfrac{a - \mu_2}{\hbar}, \ \ N_- = \dfrac{-a-\mu_4}{\hbar}
\label{SelbergParameters3}
\vspace{2ex}
\end{align}
\smallskip\\
After one substitutes (\ref{CharacterNekrasov}) into (\ref{NekFull}) and takes the
sum over $A,B$, the characters re-combine in the way that precisely reproduces the
Dotsenko-Fateev integral (\ref{Fateev-Dotsenko}), where the above two Selberg
ensembles correspond to the two groups of variables $z_i$ with $i \leq N_1$ and
$i > N_1$, respectively ($N_\pm \equiv N_{1,2}$). Thus, the AGT relation for $\beta = 1$ is derived through the character expansion of the Dotsenko-Fateev integral, and can be interpreted as duality (as illustrated in Figure 4).

It is tempting to generalize identity (\ref{CharacterNekrasov}) to $\beta \neq 1$, i.e.
to $\epsilon_1 + \epsilon_2 \neq 0$. Naively, one just has to substitute the
Schur polynomials by the Jack polynomials
{\fontsize{10pt}{0pt}
\begin{align}
J_{1}(p) = p_{1}
\label{Jack1}
\end{align}
\vspace{-2ex}
\begin{align}
J_{2}(p) = \dfrac{p_2 + \beta p_{11}}{\beta + 1}, \ \ \
J_{11}(p) = \dfrac{p_1^2}{2}-\dfrac{p_2}{2}
\label{Jack2}
\end{align}
\vspace{-2ex}
\begin{align}
J_{3}(p) = \dfrac{2 p_3 + 3 \beta p_{1} p_2 + \beta^2 p_{1}^3}{(\beta + 1)(\beta + 2)},
\
J_{21}(p) = \dfrac{(1-\beta) p_{1} p_2 - p_3 + \beta p_{1}^3 }{(\beta + 1)(\beta + 2)}, \
J_{111}(p) = \dfrac{p_3}{3} - \dfrac{p_1p_2}{2} + \frac{p_1^3}{6}
\label{Jack3}
\end{align}}
\smallskip\\
and change the power $\beta$ of the Van-der-Monde determinant

\vspace{-1ex}
\begin{align}
\prod\limits_{i<j} (z_i - z_j)^{2} \ \ \ \mapsto \ \ \ \prod\limits_{i<j} (z_i - z_j)^{2\beta}
\vspace{3ex}
\end{align}
in the definition of the Selberg averages. However, this naive $\beta$-deformation
fails to reproduce the Nekrasov coefficients $N_{AB}\big( \epsilon_1, \epsilon_2 \big)$.
In our opinion, the basic reason for the discrepancy is that $N_{AB}$ (if considered
as a rational function of $a$) has a very special structure of poles, which
accidentally coincides with that of Selberg integrals at $\beta = 1$, but for generic
$\beta$ is not captured by the Selberg integrals. To reproduce
$N_{AB}\big( \epsilon_1, \epsilon_2 \big)$ for generic
$\epsilon_1, \epsilon_2$, some \emph{clever} deformation of the r.h.s. of
(\ref{CharacterNekrasov}) is required.
\textbf{Clarifying this point would complete the direct proof of the AGT conjecture}.

\paragraph{} This paper is organized as follows.

\paragraph{Section 2} is devoted to the simple case of the AGT relation for pure
$SU(2)$. Consideration of this simple case helps to elucidate some of the important
details of the story. We describe the conformal block (as the Dotsenko-Fateev integral),
the Nekrasov functions (as explicit sums over Young diagrams) and state the AGT
relation between them. Then, using a pure gauge version of the pair-correlator
identity (\ref{CharacterNekrasov}), we derive the $\epsilon_1+\epsilon_2 = 0$
Nekrasov function from the $\beta = 1$ Dotsenko-Fateev integral.

\paragraph{Section 3} similarly deals with the AGT relation for $SU(2)$ with four
fundamental matter hypermultiplets. We describe, with the help of the pair-correlator
identity (\ref{CharacterNekrasov}), the analytical proof of equality between
the $\epsilon_1+\epsilon_2 = 0$ Nekrasov function and the $\beta = 1$ Dotsenko-Fateev
integral, for arbitrary values of masses.

\paragraph{Section 4} is devoted to analysis of the problems, which arise when one attempts to generalize our construction to generic $\beta$.

\paragraph{Section 5} is the Conclusion.

\paragraph{The Appendix} is a list of various known factorizable 1-character (Jack) and
2-character (Jack) averages in the Selberg and BGW matrix model ($\beta$-ensemble) theories,
for $\beta = 1$ and $\beta \neq 1$, which can also play a role in the future investigations.

\section{The case of pure $SU(2)$}

\subsection{Nekrasov function}

The Nekrasov sum over partitions in this case has the form

\begin{align}
Z^{{\rm pure}}_{Nek}(\Lambda) \ = \ & \sum\limits_{A,B} \ N^{{\rm pure}}_{A,B} \ \Lambda^{4|A| + 4|B|}, \ \ \ \ N^{{\rm pure}}_{A,B} = \dfrac{(\epsilon_1\epsilon_2)^{2|A|+2|B|}}{g_{A,A}(0) g_{A,B}(2a) g_{B,A}(-2a) g_{B,B}(0)}
\label{NekPure}
\end{align}
\smallskip\\
where $g_{A,B}$ denote the contributions of gauge fields into the Nekrasov function

\begin{align}
\hspace{-5ex} g_{A,B}(x) = \prod\limits_{(i,j) \in A} \Big[ x + \epsilon_1 {\rm Arm}_A(i,j) - \epsilon_2 {\rm Leg}_B(i,j) + \epsilon_1 \Big] \Big[ x + \epsilon_1 {\rm Arm}_A(i,j) - \epsilon_2 {\rm Leg}_B(i,j) - \epsilon_2 \Big]
\label{GaugeContrib}
\vspace{-2ex}
\end{align}
\smallskip\\
which has a characteristic form of a product over all the cells of the Young diagram.
For the arbitrary Young diagram $Y$, the symbols ${\rm Arm}_Y(i,j)$ and ${\rm Leg}_Y(i,j)$
denote the arm-length and leg-length of the cell $(i,j)$ in the diagram $Y$. Algebraically,
these lengths are given by the expressions
\begin{wrapfigure}{l}{130pt}
  \begin{center}
\vspace{-5ex} \includegraphics[width=120pt,height=100pt]{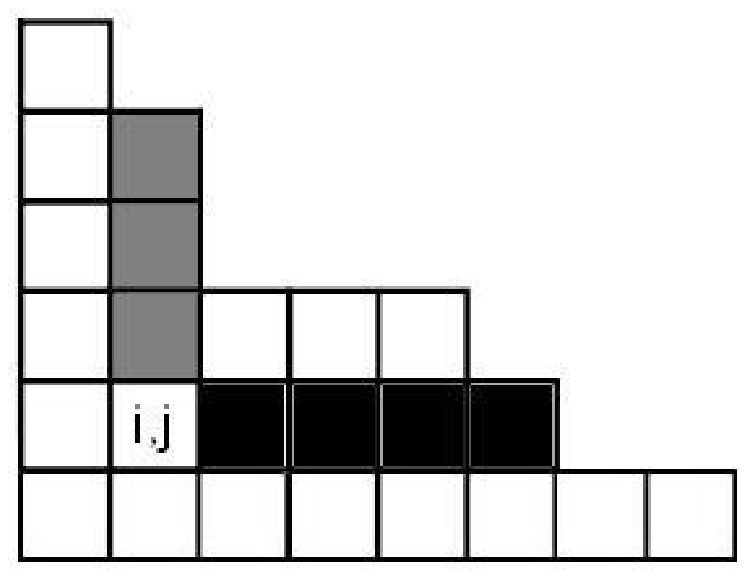}
  \end{center}
\vspace{-6ex}
  \caption{For the cell $(i,j)$ = $(2,2)$, the arm- and leg-length are shown in black and grey,
  respectively. Note that the cell can lie beyond the diagram.}
\end{wrapfigure}
\vspace{-2ex}
\begin{align}
{\rm Arm}_Y(i,j) = Y^{\prime}_j - i, \ \ \ {\rm Leg}_Y(i,j) = Y_i - j
\end{align}
where $Y^{\prime}$ stands for the transposed Young diagram.
This algebraic definition is not quite transparent: more enlightening may be
the graphical meaning of these quantities, which is shown at Figure 5.
Several first Nekrasov coefficients for pure $SU(2)$ have the form
\begin{align}
N_{[1][]} = \frac{-\epsilon_1\epsilon_2}{2a(2a+\epsilon_1+\epsilon_2)}, \ \ N_{[][1]} =
\frac{-\epsilon_1\epsilon_2}
{2a(2a-\epsilon_1-\epsilon_2)}
\label{NekPure1}
\end{align}
\vspace{-3ex}
\begin{align}
N_{[1][1]} = \frac{\epsilon^2_1\epsilon_2^2}{2! (\epsilon_1-\epsilon_2)}\cdot
\frac{2(\epsilon_1- \epsilon_2) }
{(4a^2-\epsilon_1^2)(4a^2-\epsilon_2^2)}
\label{NekPure2}
\end{align}
\vspace{-3ex}
\be
N_{[2][]} = \frac{\epsilon^2_1\epsilon_2^2}{2! (\epsilon_1-\epsilon_2)}\cdot
\frac{\epsilon_1}
{2a(2a+\epsilon_2)(2a+\epsilon_1+\epsilon_2)(2a+\epsilon_1+2\epsilon_2)},\label{NekPure3}\\
N_{[][2]} = \frac{\epsilon^2_1\epsilon_2^2}{2!(\epsilon_1-\epsilon_2)}\cdot
\frac{\epsilon_1}
{2a(2a-\epsilon_2)(2a-\epsilon_1-\epsilon_2)(2a-\epsilon_1-2\epsilon_2)},\label{NekPure4}\\
N_{[11][]} = \frac{\epsilon^2_1\epsilon_2^2}{2!(\epsilon_1-\epsilon_2)}\cdot
\frac{-\epsilon_2}
{2a(2a+\epsilon_1)(2a+\epsilon_1+\epsilon_2)(2a+2\epsilon_1+\epsilon_2)}, \label{NekPure5} \\
N_{[][11]} = -\frac{\epsilon^2_1\epsilon_2^2}{2!(\epsilon_1-\epsilon_2)}\cdot
\frac{-\epsilon_2}
{2a(2a-\epsilon_1)(2a-\epsilon_1-\epsilon_2)(2a-2\epsilon_1-\epsilon_2)}
\label{NekPure6}
\ee
Comparing with eqs.(\ref{NekMasses1})-(\ref{NekMasses6}), one can see that the case of pure
$SU(2)$ can be obtained from the more general case of $SU(2)$ with four fundamental
matter hypermultiplets by a particular pure gauge limit (PGL) (note that various
$\epsilon_{1,2}$-dependent factors emerging in the Nekrasov functions in PGL are
completely determined by the way one takes this limit):
\begin{align}
\mu_1, \mu_2, \mu_3, \mu_4 \rightarrow \infty, \ \ \ \ q \cdot \dfrac{\mu_1 \mu_2}{\epsilon_1\epsilon_2} \dfrac{\mu_3 \mu_4}{\epsilon_1 \epsilon_2} = \Lambda^4 = {\rm fixed}
\label{PGL}
\end{align}
As one can see, in this limit the Nekrasov functions get simplified. The conformal block in this
limit is also simplified \cite{nonconf,PGL} and coincides with the PGL of the 1-point toric
block \cite{DFhg}.

\subsection{Dotsenko-Fateev integral}

As explained in \cite{PGL}, the relevant Dotsenko-Fateev integral can be obtained by taking the
PGL of the initial integral (\ref{Fateev-Dotsenko}). The result is somewhat non-trivial
\cite{PGL}:

\begin{align}
\boxed{
Z^{{\rm pure}}_{DF}(\Lambda) = \Big< \ \Big< \ \det \big( 1 - \Lambda^4 U \otimes {\widetilde U} \big)^{2\beta} \ \Big>^{BGW}_+ \ \Big>^{BGW}_-
\label{DFpure}}
\end{align}
\smallskip\\
where the averaging goes over two independent $\beta$-ensembles (labeled with the symbols $+$, $-$)

\begin{align}
\Big< f(U) \Big>^{BGW}_+ = \int\limits_{n_+ \times n_+} \dfrac{[dU]_\beta}{{\rm Vol}_\beta(n_+)} \ f(U) \ Z_{BGW}\Big( n_+ + \delta \Big| t_k \Big), \ \ \ t_k = \tr (U^{+})^k/k
\end{align}
\begin{align}
\Big< f({\widetilde U}) \Big>^{BGW}_- = \int\limits_{n_- \times n_-} \dfrac{[d{\widetilde U}]_\beta}{{\rm Vol}_\beta(n_-)} \ f({\widetilde U}) \ Z_{BGW}\Big( n_- + \delta \Big| {\widetilde t}_k \Big), \ \ \ {\widetilde t}_k = \tr ({\widetilde U}^{+})^k/k
\end{align}
\smallskip\\
with the $\beta$-deformed Brezin-Gross-Witten (BGW) partition function
\cite{BGW} in the role of integrand

\begin{align}\label{PGLa}
Z_{BGW}\Big( n \Big| t_k = \tr \Psi^k / k \Big) = \int\limits_{n \times n} \dfrac{[dU]_\beta}{{\rm Vol}_\beta(n)} e^{\beta\big( \tr U^{+} + \tr (U \Psi) \big)}
\end{align}
\smallskip\\
and with $\delta = (\beta-1)/\beta$. It is checked in the same paper \cite{PGL} that
Dotsenko-Fateev integral (\ref{DFpure}) reproduces correctly the first terms of the
$\Lambda$-expansion of the conformal block.

\subsection{The AGT conjecture}

The AGT conjecture states that
\begin{align}
Z^{{\rm pure}}_{DF}(\Lambda) \ = \ Z^{{\rm pure}}_{Nek}(\Lambda)
\label{AGTpure}
\end{align}
under the following identification of parameters:
\begin{align}
& \beta = \dfrac{- \epsilon_1}{\epsilon_2}, \ n_+ = \dfrac{2a}{\epsilon_1}, \ n_- = \dfrac{-2a}{\epsilon_1}
\end{align}
Let us prove this statement in the case of $\beta = 1$.

\subsection{Proof of (\ref{AGTpure}) at $\beta = 1$}

We start from rewriting the determinant in eq.(\ref{DFpure}) in the exponential form:

\begin{align}
\det \big( 1 - \Lambda^4 U \otimes {\widetilde U} \big)^{2\beta} = \exp\left( -2\beta \sum\limits_{k=1}^{\infty} \dfrac{\Lambda^{4k}}{k} \tr U^k \tr {\widetilde U}^k \right)
\end{align}
\smallskip\\
Therefore, the Dotsenko-Fateev integral takes the form

\begin{align}
Z^{{\rm pure}}_{DF}(\Lambda) = \left< \ \left< \ \exp\left( -2\beta \sum\limits_{k=1}^{\infty} \dfrac{\Lambda^{4k}}{k} \tr U^k \tr {\widetilde U}^k \right)  \ \right>^{BGW}_+ \ \right>^{BGW}_-
\label{pureDFexp}
\end{align}
\smallskip\\
To expand this expression in characters, one can use the standard Cauchy-Stanley identity

\begin{align}\label{40}
\exp\left( \beta \sum\limits_{k=1}^{\infty} k t_k t^{\prime}_k \right) = \sum\limits_{R} j_R(t) j_R(t^{\prime})
\end{align}
\smallskip\\
where the sum is taken over all Young diagrams $R$ and $j_R$ are the normalized Jack polynomials,
$j_R = J_R / ||J_R||$ (see the Appendix for the details) which at $\beta = 1$ coincide with
the ordinary Schur polynomials:

\begin{align}
j_R\Big|_{\beta=1}=\chi_R
\end{align}
\smallskip\\
The exponent in (\ref{pureDFexp}) contains $-2\beta$ instead of $+\beta$; thus
(\ref{40}) is not directly applicable, instead one can use a trick: rewrite
it in the following form:

\begin{align}
\nonumber Z^{{\rm pure}}_{DF}(\Lambda) \ = \ & \Big< \ \Big< \ \exp\left( -2\beta \sum\limits_{k=1}^{\infty} \dfrac{\Lambda^{4k}}{k} \tr U^k \tr {\widetilde U}^k \right)  \ \Big>^{BGW}_+ \
\Big>^{BGW}_- =
\\ \nonumber & \\ \nonumber & \hspace{-10ex} = \Big< \ \Big< \exp\left( \beta \sum\limits_{k=1}^{\infty} \dfrac{\Lambda^{4k}}{k} \big( - \tr U^k \big) \tr {\widetilde U}^k \right) \exp\left( \beta \sum\limits_{k=1}^{\infty} \dfrac{\Lambda^{4k}}{k} \tr U^k \big( - \tr {\widetilde U}^k \big) \right) \Big>^{BGW}_+ \
\Big>^{BGW}_- = \\
\nonumber & \\
& \hspace{-10ex} = \sum\limits_{A,B} \ \Lambda^{4|A|+4|B|} \ \Big< \ j_A\big( - \tr U^k \big) j_B\big( \tr U^k \big) \ \Big>^{BGW}_+ \
\Big< \ j_A\big( \tr {\widetilde U}^k \big) j_B\big( - \tr {\widetilde U}^k \big)  \Big>^{BGW}_-
\label{pureDFchar}
\end{align}
\smallskip\\
At $\beta = 1$, the r.h.s. is precisely the Nekrasov function, due to the pair-correlator
identity (\ref{AppendixEq2}):

\begin{align}
\boxed{
\Big< \chi_{A}\big( - \tr U^k \big) \ \chi_B\big(\tr U^k\big) \Big>^{BGW}_+ \ \ \Big< \chi_A\big( \tr {\widetilde U}^k \big) \ \chi_{B}\big( -  \tr {\widetilde U}^k \big) \Big>^{BGW}_- = N^{{\rm pure}}_{A,B} \Big|_{\epsilon_1 + \epsilon_2 = 0}
\label{IdentityPure}}
\end{align}
\smallskip\\
which is the pure gauge limit of (\ref{CharacterNekrasov}) and is considered in more detail in
the Appendix, see eq.(\ref{AppendixEq2}). Substituting this into
(\ref{pureDFchar}), one obtains

\begin{align}
Z^{{\rm pure}}_{DF}(\Lambda)\Big|_{\beta=1} =  Z^{{\rm pure}}_{Nek}(\Lambda)\Big|_{\epsilon_1 + \epsilon_2 = 0}
\end{align}
\smallskip\\
and this completes the proof.

It may even seem that the only non-trivial part of this calculation is the pair-correlator
identity (\ref{IdentityPure}). However, the identity itself is nothing but a technical detail.
Really important is a duality: the existence of the quadrilinear character expansion
(\ref{pureDFchar}). Eq. (\ref{pureDFchar}) contains both a sum over $A,B$ diagrams and an
average over $"+","-"$ ensembles, and reduces either to the Nekrasov function
(\ref{IdentityPure}) or to the Dotsenko-Fateev integral (\ref{DFpure}) if one evaluates
either the double average or the double sum, respectively. This is a typical duality, only
realized at a very simple algebraic level with the help of characters. Let us now include
masses into our consideration.

\section{The case of $SU(2)$ with 4 fundamentals}

\subsection{Nekrasov function}
The Nekrasov function for this case has the form

\begin{align}
Z_{Nek}(q) \ = \ & \sum\limits_{A,B} \ N_{A,B} \ q^{|A| + |B|}
\label{NekFullMasses}
\end{align}
\smallskip\\
with coefficients
\begin{align}
N_{A,B} = \dfrac{\prod_{k = 1}^{4} f_A(\mu_k+a)f_A(\mu_k-a)}{g_{A,A}(0) g_{A,B}(2a) g_{B,A}(-2a) g_{B,B}(0)}
\label{NekCoefMasses}
\end{align}
where in addition to contributions (\ref{GaugeContrib}) of gauge fields, one now has matter contributions:
\begin{align}
f_A(z) = \prod\limits_{(i,j) \in A} \Big[ z + \epsilon_1 (i - 1) + \epsilon_2 (j-1) \Big]
\label{MassContrib}
\end{align}
A few first Nekrasov coefficients $N_{A,B}$ are written in
eqs.(\ref{NekMasses1})-(\ref{NekMasses5}).

\subsection{Dotsenko-Fateev integral}
The Dotsenko-Fateev integral for this case has the form (\ref{Fateev-Dotsenko}), but as was noticed
a while ago \cite{ito}, for the purposes of $q$-expansion it is more convenient to rewrite this
integral (of course, omitting the $U(1)$ prefactors, which are irrelevant for comparison with the
Nekrasov functions) as a double average:

\begin{align}
Z_{DF}(q) = \left< \ \left< \ \ \prod\limits_{i = 1}^{N_+} (1 - q x_i)^{v_-} \prod\limits_{j = 1}^{N_-} (1 - q y_j)^{v_+} \prod\limits_{i = 1}^{N_+} \prod\limits_{j = 1}^{N_-} (1 - q x_i y_j)^{2\beta} \ \right>_+ \ \right>_-
\label{NekMamo}
\end{align}
\smallskip\\
where the averaging goes over two independent ensembles (labeled with symbols $+$ and $-$ )
of variables
$x_1, \ldots, x_{N_+}$ and $y_1, \ldots, y_{N_-}$ ("eigenvalues" in matrix model terms) as follows:
\begin{align}
\Big< f \Big>_+ \ = \ \dfrac{1}{S_{+}}
\int\limits_{0}^{1} dx_1 \ldots \int\limits_{0}^{1} dx_{N_+} \prod\limits_{i<j} (x_i - x_j)^{2\beta} \prod\limits_{i} x_i^{u_+} (x_i - 1)^{v_+} \ f\big(x_1, \ldots, x_{N_+}\big)
\label{xAverage}
\end{align}
\begin{align}
\Big< f \Big>_- \ = \ \dfrac{1}{S_{-}}
\int\limits_{0}^{1} dy_1 \ldots \int\limits_{0}^{1} dy_{N_-} \prod\limits_{i<j} (y_i - y_j)^{2\beta} \prod\limits_{i} y_i^{u_-} (y_i - 1)^{v_-} \ f\big(y_1, \ldots, y_{N_-}\big)
\label{yAverage}
\end{align}
with the normalization constants
\begin{align}
S_{\pm} = \int\limits_{\gamma_{\pm}} dz_1 \ldots dz_{N} \prod\limits_{i < j} (z_i - z_j)^{2\beta} \prod\limits_{i} z_i^{u_\pm} (z_i - 1)^{v_\pm}
\end{align}
needed to satisfy $\Big< 1 \Big>_+ = \Big< 1 \Big>_- = 1$.

\subsection{The AGT conjecture}
The AGT conjecture states that
\begin{align}
Z_{DF}(q) \ = \ Z_{Nek}(q)
\label{AGTmasses}
\end{align}
under the following identification of parameters:
\begin{align}
& \beta = \dfrac{- \epsilon_1}{\epsilon_2}, \ N_+ = \dfrac{a - \mu_2}{\epsilon_1}, \ N_- = \dfrac{- a - \mu_4}{\epsilon_1}
\label{mAGT1}
 \\
& \nonumber \\
& u_+ = \dfrac{\mu_1 - \mu_2 - \epsilon_1 - \epsilon_2}{\epsilon_2}, \ u_- = \dfrac{\mu_3 - \mu_4 - \epsilon_1 - \epsilon_2}{\epsilon_2} \label{mAGT2}\\
& \nonumber \\
& v_+ = \dfrac{-\mu_1-\mu_2}{\epsilon_2}, \ v_- = \dfrac{-\mu_3-\mu_4}{\epsilon_2} \label{mAGT3}
\end{align}
Let us prove this statement in the case of $\beta = 1$.

\subsection{Proof at $\beta = 1$}

The proof goes completely similar to the BGW case. Likewise, we start from rewriting the
Dotsenko-Fateev integrand in an exponential form, and then use the Cauchy-Stanley identity to
perform an expansion in the basis of Schur/Jack symmetric polynomials:

\begin{align}
\prod\limits_{i = 1}^{N_+} (1 - q x_i)^{v_-} \prod\limits_{j = 1}^{N_-} (1 - q y_j)^{v_+} \prod\limits_{i = 1}^{N_+} \prod\limits_{j = 1}^{N_-} (1 - q x_i y_j)^{2\beta} = \emph{}\nn
\end{align}
\begin{align}
\emph{} = \exp\left( - \beta \sum\limits_{k=1}^{\infty} \dfrac{q^k}{k} {\widetilde p}_k( p_k + v_+) \right) \exp\left( - \beta \sum\limits_{k=1}^{\infty} \dfrac{q^k}{k} p_k( {\widetilde p}_k + v_-) \right) =
\end{align}
\begin{align}
\emph{} = \sum\limits_{A,B} q^{|A|+|B|} j_{A}\big( - p_k - v_+ \big) j_B\big(p_k\big) j_A\big( {\widetilde p}_k \big) j_{B}\big( -  {\widetilde p}_k - v_- \big)\nn
\label{MassesExpansion}
\end{align}
\smallskip\\
where $p_k = \sum_i x_i^k$ and ${\widetilde p}_k = \sum_i y_i^k$. Therefore, the
Dotsenko-Fateev integral takes the form

\begin{align}
Z_{DF}(\Lambda) = \sum\limits_{A,B} \ q^{|A|+|B|} \ \Big< j_{A}\big( - p_k - v_+ \big) \ j_B\big(p_k\big) \Big>_+ \ \Big< j_A\big( {\widetilde p}_k \big) \ j_{B}\big( -  {\widetilde p}_k - v_- \big) \Big>_-
\end{align}
\smallskip\\
At $\beta = 1$, the correlators at the r.h.s. precisely reproduce the Nekrasov function
(see (\ref{AppendixEq1})):

\begin{align}
\Big< \chi_{A}\big( - p_k - v_+ \big) \ \chi_B\big(p_k\big) \Big>_+ \ \Big< \chi_A\big( {\widetilde p}_k \big) \ \chi_{B}\big( -  {\widetilde p}_k - v_- \big) \Big>_- = N_{A,B} \Big|_{\epsilon_1 + \epsilon_2 = 0}
\label{IdentityMasses}
\end{align}
\smallskip\\
Substituting this into (\ref{MassesExpansion}), one obtains

\begin{align}
Z_{DF}(\Lambda)\Big|_{\beta=1} =  Z_{Nek}(\Lambda)\Big|_{\epsilon_1 + \epsilon_2 = 0}
\end{align}
\smallskip\\
which completes the proof.

\section{Problems with generalization to $\beta \neq 1$}

\begin{wrapfigure}{r}{130pt}
  \begin{center}
\vspace{-5ex} \includegraphics[width=120pt,height=100pt]{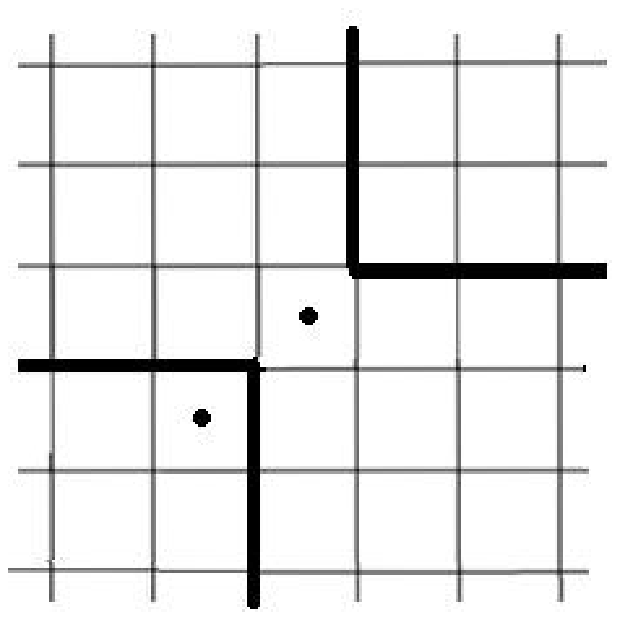}
  \end{center}
\vspace{-4ex}
  \caption{Poles of $N_{[1][]}(z)$.}
\end{wrapfigure}

The basic puzzle of the AGT relation for $\beta\neq 1$
is a different structure of poles at the two sides
of the equality. The conformal block has poles at zeroes of the Kac determinant,
i.e. at $z=m\epsilon_1 + n\epsilon_2$ with $mn>0$,
while the poles of the particular Nekrasov functions
$N_{AB}(z)$   (here $z=2a$)
occur also at $mn\leq 0$.
Transition from the conformal blocks to the Selberg or BGW
pair correlators of characters, exploited in the present paper, does not help: their poles are
still at $mn>0$, just as for the conformal blocks.

In this section, Figures 6-11 are used to illustrate the issue of poles. In these pictures, the
square lattice
represents the set of possible linear combinations $m\epsilon_1 + n\epsilon_2$,
dots represent positions of poles, and the bold area in the top right corner (and its
mirror image in the bottom left corner) represents the part of the lattice with $mn>0$,
where zeroes of the Kac determinant may be situated. The horizontal and vertical directions
correspond to $\epsilon_1$ and $\epsilon_2$, respectively. The central cell of the lattice
corresponds
to the point $(m,n) = (0,0)$.

\begin{wrapfigure}{r}{130pt}
  \begin{center}
\vspace{-5ex} \includegraphics[width=120pt,height=100pt]{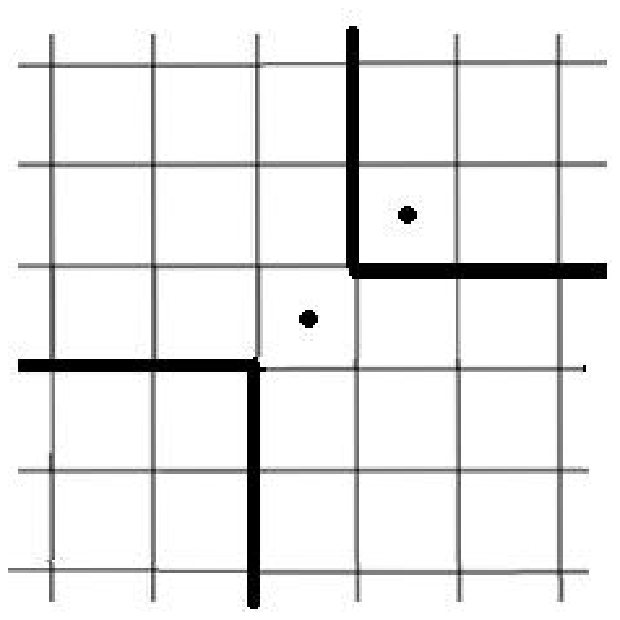}
  \end{center}
\vspace{-4ex}
  \caption{Poles of $N_{[][1]}(z)$.}
\end{wrapfigure}
Because of the problem of poles, it is unclear if it is at all possible
to extend the relations like (\ref{CharacterNekrasov}) and (\ref{IdentityPure})
to $\beta\neq 1$.
What happens at $\beta=1$ is that only the
difference $m-n$ matters, and all the poles can be
projected from the plane to a single line
$z = (m-n)\epsilon_1$, and the difference between
the sets with $mn > 0$ and $mn\leq 0$ disappears.
This phenomenon at $\epsilon_1+\epsilon_2=0$ is illustrated in Figure 8.

Of course, the extra poles of the particular
Nekrasov coefficients $N_{AB}(z)$ drop away from
their sum, i.e. from the LMNS partition
function, which is AGT-related to the conformal block.
Thus the real puzzle is,
what at all is the real role of the individual
$N_{AB}$, i.e. why does the linear basis with
the nicely factorizable coefficients
(as functions of $\mu$'s and $\epsilon$'s)
include extra poles in the $z$-variable.
\begin{wrapfigure}{l}{130pt}
  \begin{center}
\vspace{-5ex} \includegraphics[width=120pt,height=100pt]{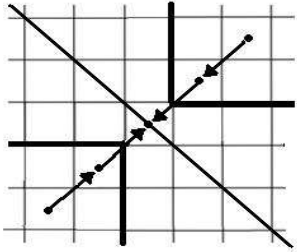}
  \end{center}
\vspace{-4ex}
  \caption{At $\epsilon_1+\epsilon_2=0$, all poles with equal $m-n$ become indistinguishable.}
\end{wrapfigure}
Anyhow, if $N_{AB}(z)$ are relevant,
their Selberg or BGW interpretation is still missed
when $\beta\neq 1$.

In what follows we illustrate the problem
at the first two levels of the Young diagram
expansion. For this, in addition to explicit formulas
for the Nekrasov functions in (\ref{NekMasses1})-(\ref{NekMasses6})
and (\ref{NekPure1})-(\ref{NekPure6}),
one also needs explicit formulas for the pair
correlators of the $\beta$-characters
(i.e. the normalized Jack polynomials, see the Appendix).
These are listed in Tables 2-4 and 7-9 for the BGW case.
Actually we need only the entries of Table 3,
the other two are added to illustrate factorizability
properties and to provide some data
for the future study of alternatives to eq.(\ref{CharacterNekrasov}):
clearly, instead of
\be
R_{AB}(z)R_{BA}(-z) =<j_A(p)j_B(-p)>_+
<j_B(p)j_A(-p)>_-
\ee
\begin{wrapfigure}{l}{110pt}
  \begin{center}
\vspace{-4ex} \includegraphics[width=110pt,height=90pt]{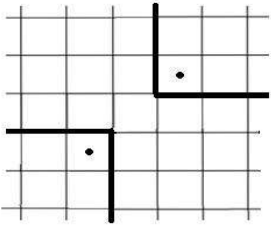}
  \end{center}
\vspace{-4ex}
  \caption{Poles of the sum $N_{[1][]}+N_{[][1]}$. The "extra" poles vanish.}
\end{wrapfigure}
one could also use another types of correlators:
\be
S_{AB}(z)Q_{AB}(-z) = <j_A(p)j_B(p)>_+
<j_A(-p)j_B(-p)>_-
\ee
or
\be
Q_{AB}(z)S_{AB}(-z) = <j_A(-p)j_B(-p)>_+
<j_A(p)j_B(p)>_-
\ee
or any linear combination of the three. Tables 2 and 7 are devoted to correlators $S_{AB}$,
tables 3 and 8 to $R_{AB}$, tables 4 and 9 to $Q_{AB}$.

Formulas for the Selberg correlators are more lengthy, but their properties are essentially
the same, see Tables 5 and 6. Actual examples below are given for the simpler BGW case, i.e.
relevant for the AGT relation in the pure gauge limit.

\begin{figure}[t]
\begin{center}
\includegraphics[width=80pt]{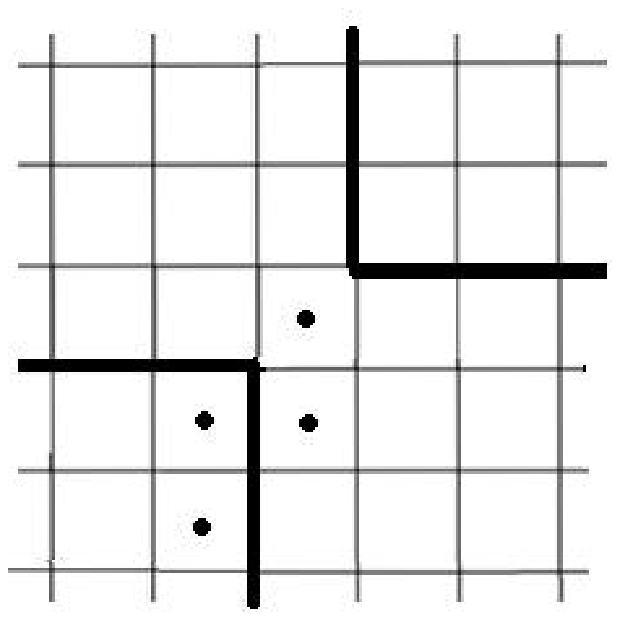}
\ \ \includegraphics[width=80pt]{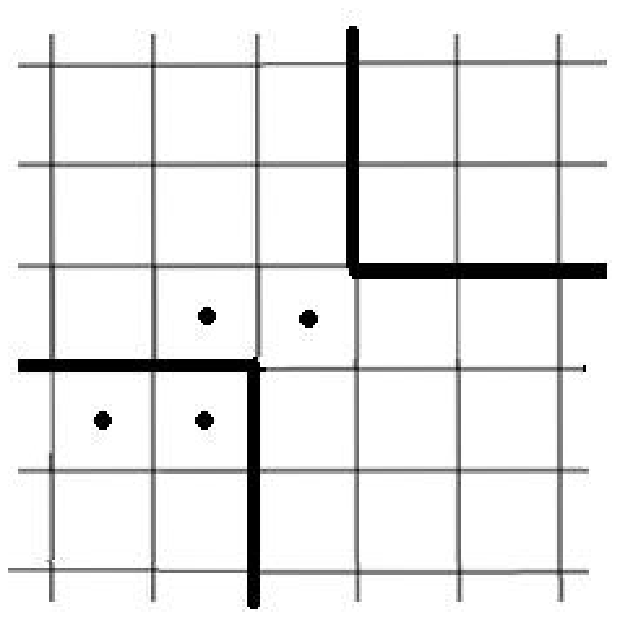}
\ \ \includegraphics[width=80pt]{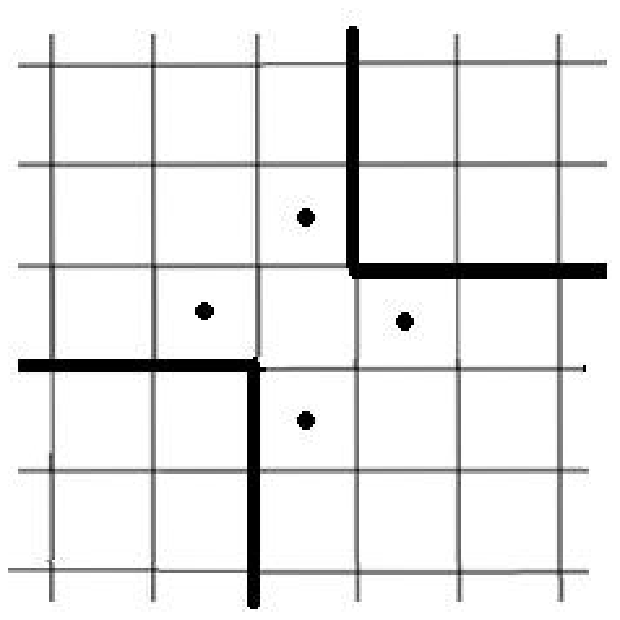}
\ \ \includegraphics[width=80pt]{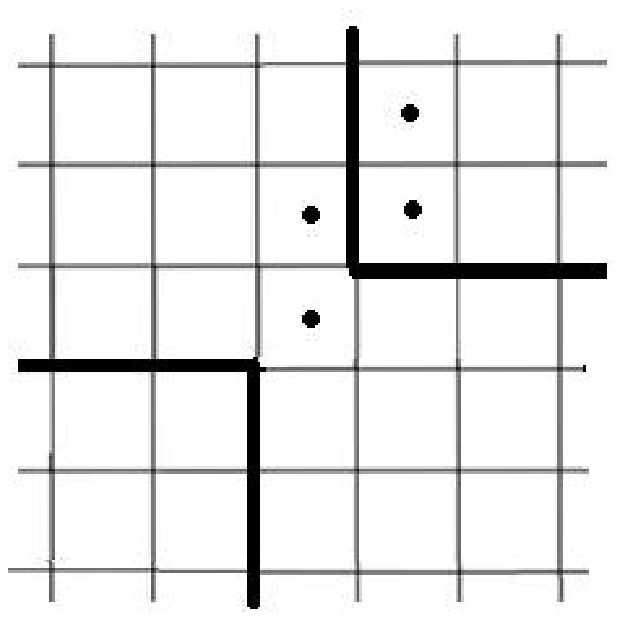}
\ \ \includegraphics[width=80pt]{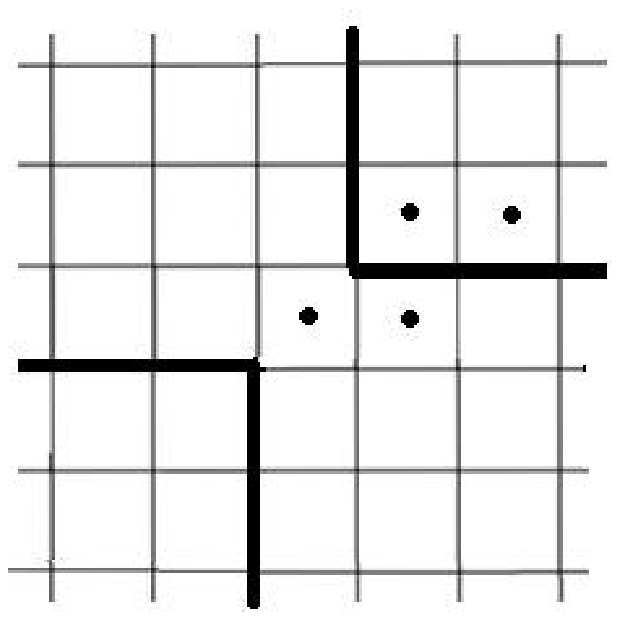}
\end{center}
\vspace{-4ex}
\caption{Poles of $N_{[2][]},N_{[11][]},N_{[1][1]},N_{[][2]}$ and $N_{[][11]}$, respectively.}
\end{figure}

\paragraph{Level 1.}
At level one, the relation looks like

\be
N_{[1],[]}(z) + N_{[],[1]}(z) = -\frac{\epsilon_1\epsilon_2}{z(z-\epsilon_1-\epsilon_2)}
- \frac{\epsilon_1\epsilon_2}{z(z+\epsilon_1+\epsilon_2)} =\nn
\ee
\be
= -\frac{2\epsilon_1\epsilon_2}{(z-\epsilon_1-\epsilon_2)(z+\epsilon_1+\epsilon_2)}
= R_{[1],[]}(z)R_{[],[1]}(-z) + R_{[],[1]}(z)R_{[1],[]}(-z)
\label{RelLvl1}
\ee

The auxiliary poles, which are present at the particular Nekrasov coefficients, but disappear
from the whole sum (see Figure 10) in this case are represented by a single pole at $z = 0$.
Moral: the individual $N_{10}(z)$ and $N_{01}(z)$ can not be expressed through
$R_{10}(\pm z)$ and $R_{01}(\pm z)$, but their sum can, as shown by relation (\ref{RelLvl1}).
At $\epsilon_1 = -\epsilon_2$, however, eq. (\ref{RelLvl1}) gets simplified: the l.h.s. sum
($N + N$) and the r.h.s. sum ($RR + RR$) become equal term by term! This is illustrated by
\be
\boxed{
\begin{array}{c}
\frac{1}{(00)(11)} + \frac{1}{(00)(-1,-1)} = \frac{2}{(11)(-1,-1)}\\
\\ \downarrow \ \ \beta=1 \\  \\
\frac{1}{(00)^2} + \frac{1}{(00)^2} = \frac{2}{(00)^2}
\end{array}}
\ee
\begin{wrapfigure}{r}{120pt}
  \begin{center}
\vspace{-5ex} \includegraphics[width=120pt,height=100pt]{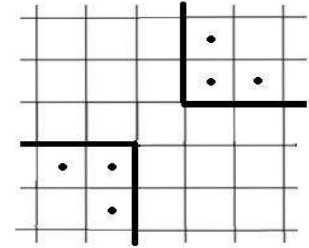}
  \end{center}
\vspace{-4ex}
  \caption{Poles of the sum $N_{[2][]}+N_{[11][]}+N_{[1][1]}+N_{[][2]}+N_{[][11]}$.
  The "extra" poles vanish.}
\vspace{-5ex}
\end{wrapfigure}
in the abbreviated notation $(m,n) = (z-m\epsilon_1-n\epsilon_2)$. The same phenomenon
of transformation of a complicated equality between the whole sums at generic
$\epsilon_1+\epsilon_2$ into the very simple equality between individual terms at
$\epsilon_1+\epsilon_2=0$ persists at higher levels.

\paragraph{}This shows that, in fact, $\epsilon_1 = -\epsilon_2$ is a highly distinguished case.
In this case, simply the passing to the basis of characters completely reveals the underlying
structure behind the AGT relation, formulated in the present paper in terms of bilinear
correlators in the Selberg models. The aim of this section is to stress that for general
$\epsilon_1,\epsilon_2$ the relation between the Nekrasov functions and Selberg correlators
is still missed and is probably more sophisticated. Finding such a relation would be crucial
for development in this research direction.

\pagebreak

\paragraph{Level 2.}At level two, the relation looks like

\begin{align}
\hspace{-2ex} \sum\limits_{|A|+|B|=2} N_{A,B}(z) = (\ref{NekPure2}) + (\ref{NekPure3}) + (\ref{NekPure4}) + (\ref{NekPure5}) + (\ref{NekPure6}) = \emph{}\nn
\end{align}
\begin{align}
\emph{} = \dfrac{\epsilon_1^2\epsilon_2^2(2 z^2-8 \epsilon_{1}^2-17 \epsilon_{1} \epsilon_{2}-8 \epsilon_{2}^2)}{(z-2 \epsilon_{1}-\epsilon_{2}) (z+\epsilon_{2}+\epsilon_{1}) (z+2 \epsilon_{2}+\epsilon_{1}) (z+2 \epsilon_{1}+\epsilon_{2}) (z-2 \epsilon_{2}-\epsilon_{1}) (z-\epsilon_{2}-\epsilon_{1})} =
\end{align}
\begin{align}
\emph{} = R_{[1],[1]}(z)R_{[1],[1]}(-z) + R_{[2],[]}(z)R_{[],[2]}(-z) + R_{[],[2]}(z)R_{[2],[]}(-z) +\emph{}\nn
\end{align}
\begin{align}
\emph{} +  R_{[1,1],[]}(z)R_{[],[1,1]}(-z) + R_{[],[1,1]}(z)R_{[1,1],[]}(-z)\nn
\end{align}
Again, this is a complicated relation, not quite expectable if one takes a look simply at the
rational functions at the l.h.s. and the r.h.s.: this time, 5 auxiliary poles at $z = 0$,
$ \epsilon_1,\epsilon_2$, $-\epsilon_1$ and $-\epsilon_2$ disappear from the final sum
(as illustrated in Figure 11). In analogy with the level 1 case, at
$\epsilon_1+\epsilon_2=0$ the relation gets satisfied term by term and thus completely
transparent:

{\fontsize{7pt}{0pt}
\be
\hspace{-2ex}\frac{\epsilon_1}{(00)(01)(11)(12)}
+ \frac{\epsilon_1}{(00)(0-1)(-1,-1)(-1,-2)}
+ \frac{2\epsilon_{12}}{(01)(0,-1)(1,0)(-1,0)}
- \frac{\epsilon_2}{(00)(10)(11)(21)}
-\frac{\epsilon_2}{(00)(-1,0)(-1,-1)(-2,-1)}\nn
\ee
\be
=
\frac{\epsilon_1(23)(-2,-1)+\epsilon_1(21)(-2,-3)
+ 2\epsilon_{12}(22)(-2,-2)
-\epsilon_2(32)(-1,-2)-\epsilon_2(1,2)(-3,-2)}
{(11)(12)(21)(-1,-1)(-1,-2)(-2,-1)}
\ee
$$\downarrow \beta=1$$
\be
\frac{1}{(00)^2(01)^2} + \frac{1}{(00)^2(10)^2}
+ \frac{4}{(01)^2(10)^2}
+ \frac{1}{(00)^2(10)^2} + \frac{1}{(00)^2(01)^2}
=
\frac{(01)^2 + (10)^2 + 4(00)^2 + (10)^2 + (01)^2}
{(00)^2(01)^2(10)^2}\nn
\ee
}
Of course, instead of the combinations of correlators $R(z)R(-z)$ one could also use
the combinations $S(z)Q(-z)$ or $Q(z)S(-z)$ or some linear combination like
$S(z)Q(-z)+Q(z)S(-z)$. All these formulations are equivalent: each time there is a
transcendental equality of sums of rational functions, which at $\epsilon_1+\epsilon_2=0$
turns into a term-by-term equality.

\paragraph{Higher levels.} At higher levels, things get even more sophisticated. A new feature,
which appears at this level of consideration, is that only the correlators $S_{AB}$ remain
factorized, while the correlators $R_{AB}$ and $Q_{AB}$ at $\beta \neq 1$ contain
non-factorizable expressions in numerators. Thus, it becomes impossible to illustrate the
phenomenon by using the shorthand notation $(n,m)$. However, the phenomenon itself does not
change: at $\epsilon_1+\epsilon_2=0$, the relation between the Nekrasov side and
the Selberg side becomes termwise.

\pagebreak

\section{Conclusion}

\begin{figure}\label{dual}
\unitlength 0.5mm 
\linethickness{0.4pt}
\ifx\plotpoint\undefined\newsavebox{\plotpoint}\fi 
\begin{picture}(00,173.5)(-15,-20)
\put(129.5,129.25){\line(1,1){19}}
\put(148.5,148.5){\line(-1,1){22}}
\put(153.5,148.5){\line(1,1){22}}
\put(153.5,148.5){\line(1,-1){19}}
\multiput(150,151)(.664063,.664063){33}{{\rule{.4pt}{.4pt}}}
\multiput(150,151)(-.647059,.647059){35}{{\rule{.4pt}{.4pt}}}
\multiput(151.18,145.68)(-.65179,-.65179){29}{{\rule{.4pt}{.4pt}}}
\multiput(151.18,145.68)(.64286,-.66964){29}{{\rule{.4pt}{.4pt}}}
\put(124.25,118){\vector(-2,-3){15}}
\put(30,89.75){\line(1,-1){19.25}}
\put(49,70.5){\line(-1,-1){17.5}}
\multiput(32.68,92.43)(.66667,-.65){31}{{\rule{.4pt}{.4pt}}}
\multiput(52.68,72.93)(.98913,0){47}{{\rule{.4pt}{.4pt}}}
\multiput(35.18,50.93)(.65385,.67308){27}{{\rule{.4pt}{.4pt}}}
\multiput(52.18,68.43)(.984043,-.005319){48}{{\rule{.4pt}{.4pt}}}
\put(101.75,70.75){\line(1,1){20}}
\put(102,70.5){\line(1,-1){19}}
\multiput(98.43,67.93)(.65179,-.67857){29}{{\rule{.4pt}{.4pt}}}
\multiput(98.68,73.18)(.65,.66667){31}{{\rule{.4pt}{.4pt}}}
\put(199.25,93.5){\line(1,-1){17}}
\put(216.25,76.5){\line(0,-1){39}}
\put(216.25,37.5){\line(-1,-1){17.75}}
\put(220.5,76.5){\line(1,1){19.5}}
\put(220.5,76.25){\line(0,-1){39.25}}
\put(220.5,37){\line(1,-1){17.5}}
\multiput(202.43,96.93)(.65,-.66){26}{{\rule{.4pt}{.4pt}}}
\multiput(218.68,80.43)(.64286,.65179){29}{{\rule{.4pt}{.4pt}}}
\multiput(201.68,17.93)(.65,.65){26}{{\rule{.4pt}{.4pt}}}
\multiput(217.93,34.18)(.67308,-.65385){27}{{\rule{.4pt}{.4pt}}}
\put(178.5,122){\vector(1,-1){15}}
\put(0,-10){\makebox(0,0)[cc]{$\sum_{A,B}N_{AB}$}}
\put(23,-10){\makebox(0,0)[cc]{=}}
\put(93,-10){\makebox(0,0)[cc]
{\hspace{-6ex}$\overbrace{\sum_{AB}\int_z \chi_A(z)\chi_B(z)\int_y \chi_A(y)\chi_B(y)}$\hspace{-2ex}}}
\put(152,-10){\makebox(0,0)[cc]{\hspace{-2ex}=}}
\put(233,-10){\makebox(0,0)[cc]
{\hspace{-12ex}$\overbrace{\int_{z,y} \sum_A \chi_A(z)\chi_A(y)\sum_B  \chi_B(z)\chi_B(y)}$\hspace{-2ex}}}
\put(285,-10){\makebox(0,0)[cc]{=}}
\put(297,-10){\makebox(0,0)[cc]{${\cal B}(q)$}}
\put(74,38.75){\vector(0,-1){15}}
\put(216,18.75){\vector(0,-1){15}}
\end{picture}
\caption{The picture of Nekrasov functions/conformal block duality expressed by the
Hubbard-Stratonovich type formula (\ref{HS}). The symbol $\int_z$ here denotes integration with the Selberg measure over variables $z_i$, and the symbol $\sum_A$ denotes summation over all Young diagrams $A$.}
\end{figure}
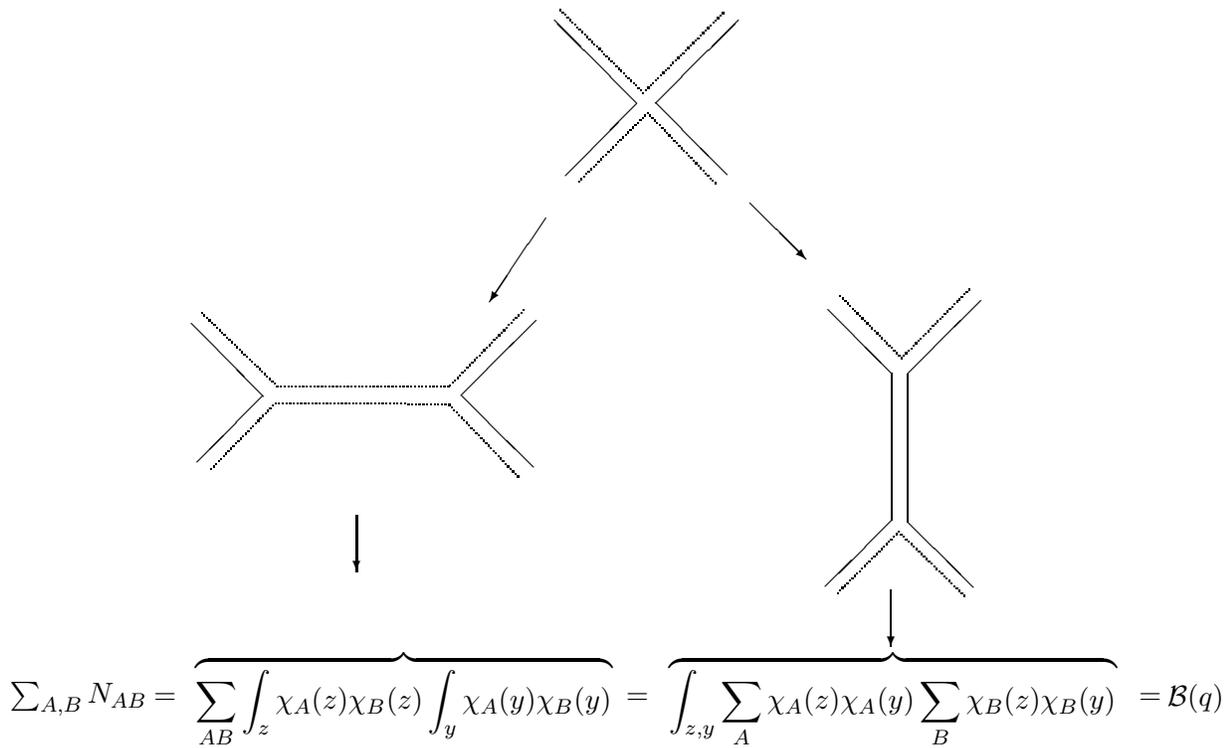

In this paper we succeeded in interpreting the AGT relation as the standard duality
relation of the Hubbard-Stratonovich type, see Fig.\ref{dual}:
\be\label{HS}
\sum_{a,b}\left(\sum_i X_i^a X_i^b\right)\left(\sum_j X_j^a X_j^b\right)
=\sum_{a,b,i,j} X_i^a X_i^b X_j^a X_j^b
= \sum_{i,j}\left(\sum_a X_i^a X_j^a\right)\left(\sum_b X_i^b X_j^b\right)
\ee
The role of $X_i^a$ is played by the $GL(\infty)$
characters $\chi_A(p)$.
This provides a very direct and conceptually clear
proof of the AGT relation,
very different from both the various formal proofs,
suggested in \cite{MMhypergeom,FLit1,Polish,Jap,AlbaLit},
and more transcendental projects like
\cite{DV,MMStalk} etc.
Moreover, as a byproduct we found a new representation
for the particular Nekrasov functions $N_{AB}$
through the pair correlators of characters in relevant
matrix models (like the Selberg or BGW ones).

Unfortunately, all this works so nicely only in the
particular case of $\beta=1$.
The extra poles puzzle at $\beta\neq 1$,
which we described above in s.4,
remains unresolved and generalization of the duality
interpretation to $\beta\neq 1$ is still missed.
No representation of the individual Nekrasov functions
$N_{AB}(z)$ in terms of pair correlators of characters
is found for $\beta\neq 1$: they simply possess more
poles than the known correlators.

At the same time, generalizations in other directions:
from 4-point to generic conformal blocks
(at least, spherical and elliptic)
and from the $U(2)$/Virasoro symmetry to
$U(N)$/$W_N$ seem straightforward.
In both cases polylinear, rather than bilinear
combinations of pair correlators are going to arise.

The technical base of our consideration is the further
generalization of the Selberg/Kadell formulas
from single to pair correlators of characters
in the Selberg and BGW models, given by eqs.(\ref{AppendixEq1})
and (\ref{AppendixEq2}) respectively.
We did not describe a proof of these formulas, it is straightforward
done within the standard approaches
(e.g., from the singularity analysis to the Ward identities).
What deserves to be mentioned, these correlators are
{\it different} from another important set of correlators
recently considered in \cite{AlbaLit} (see also the end part of the Appendix).
The two most important differences are:
(i) ours have poles, while the non-trivial part of those in \cite{AlbaLit} have only zeroes;
(ii) ours remain non-trivial in the pure gauge limit
(the Selberg correlators turn into the BGW ones),
while the non-trivial part of those in \cite{AlbaLit} becomes trivial.
An advantage of the correlators in \cite{AlbaLit} could be that they
remain factorized for $\beta\neq 1$, just like
our $R_{AB}(z)$, unfortunately, they are also not
sufficient to describe the individual
Nekrasov functions $N_{AB}(z)$ for $\beta\neq 1$.

To summarize, \textbf{the AGT relation is now clearly understood in two limits:}
\textbf{for $c=\infty$} \cite{MMhypergeom}, when conformal blocks become ordinary
hypergeometric series, and \textbf{for $c=1$} when they possess the free fermion
representation and, as we explained in the present paper, are related to
the Nekrasov functions
by the most naive duality transformation {\it a la} (\ref{HS}).
An interpolation between these two extreme cases still remains to be found.

\section*{Acknowledgements}

Our work is partly supported by Ministry of Education and Science of
the Russian Federation under contract 02.740.11.0608, by RFBR
grant 10-01-00536-a,
by joint grants 09-02-90493-Ukr, 09-01-92440-CE, 09-02-91005-ANF,
10-02-92109-Yaf-a. The work of A.Morozov was also supported in part
by CNRS.

\pagebreak

\section*{Appendix. Averages of Jack polynomials in $\beta$-ensembles}

The present section lists various known averages of the Jack
polynomials in various $\beta$-ensembles, which possess remarkably simple
factorization properties. In this paper, we demonstrated that, in some cases, averages
of the Jack polynomials are directly related to the Nekrasov functions. Further progress in
understanding of these averages may lead to a complete reformulation of the Nekrasov functions
in terms of matrix model ($\beta$-ensemble) theory.

\subsection*{Jack polynomials}

The Jack polynomials form a distinguished basis in the space of all symmetric polynomials. They
are often used in two
different versions: the "un-normalized" $J_Y$ (see \cite[Appendix 1]{PGL} for a general
definition)
\begin{align*}
& J_{1}(p_k) = p_{1}\\&
\\&
J_{2}(p_k) =  \dfrac{p_2 + \beta p_{11}}{\beta + 1}, \ \ \ \ J_{11}(p_k) =  \dfrac{1}{2} \big( p_{1}^2 - p_2 \big) \\&
\\&
J_{3}(p_k) = \dfrac{2 p_3 + 3 \beta p_{1} p_2 + \beta^2 p_{1}^3}{(\beta + 1)(\beta + 2)}, \  J_{21}(p_k) = \dfrac{(1-\beta) p_{1} p_2 - p_3 + \beta p_{1}^3 }{(\beta + 1)(\beta + 2)}, \
J_{111}(p_k) = \dfrac{1}{6} p_{1}^3 - \dfrac{1}{2} p_{1} p_2 + \dfrac{1}{3} p_3 \\
\end{align*}
and the "normalized" $j_Y = \dfrac{J_Y}{||J_Y||}$, where $||J_Y||$ is a natural norm w.r.t. the
orthogonality

\begin{align}
\Big< J_A \Big| J_B \Big> = \delta_{AB} ||J_A||^2, \ \ \ \ \Big< p_A \Big| p_B \Big> \equiv \left. \prod_j \dfrac{B_j}{\beta} \dfrac{\partial}{\partial p_{B_j}} \prod_i p_{A_i} \right|_{p = 0}
\end{align}
\smallskip\\
Several first norms are given in the following table:

\begin{center}
\begin{tabular}{c|cc}
$A$ & $||J_A||$ & \rule{0pt}{9mm} \\
\hline
$ [1] $&$ \sqrt{\dfrac{1}{\beta}} $&$ \rule{0pt}{9mm} $\\
$ [2] $&$ \sqrt{\dfrac{2}{\beta^2(\beta+1)}} $&$ \rule{0pt}{9mm} $\\
$ [1,1] $&$ \sqrt{\dfrac{1+\beta}{2\beta}} $&$ \rule{0pt}{9mm} $\\
$ [3] $&$ \sqrt{\dfrac{6}{\beta(\beta+1)(\beta+2)}} $&$ \rule{0pt}{9mm} $\\
$ [2,1] $&$ \sqrt{\dfrac{\beta+2}{\beta(2\beta+1)}} $&$ \rule{0pt}{9mm} $\\
$ [1,1,1] $&$ \sqrt{\dfrac{(1+\beta)(1+2\beta)}{6\beta^3}} $&$ \rule{0pt}{9mm} $\\
\end{tabular}
\end{center}

\subsection*{$\beta$-ensembles}

Here we consider just two $\beta$-ensembles. The \textbf{Selberg} averaging is defined as an
integral

\begin{align}
\Big< f \Big>^{{\rm Selb}} \ = \ \dfrac{
\int\limits_{0}^{1} dz_1 \ldots \int\limits_{0}^{1} dz_{N} \prod\limits_{i<j} (z_i - z_j)^{2} \prod\limits_{i} z_i^{u} (z_i - 1)^{v} \ f\big(z_1, \ldots, z_{N}\big)}{\int\limits_{0}^{1} dz_1 \ldots \int\limits_{0}^{1} dz_{N} \prod\limits_{i<j} (z_i - z_j)^{2} \prod\limits_{i} z_i^{u} (z_i - 1)^{v} }
\end{align}
\smallskip\\
The \textbf{BGW} averaging is most simply defined as PGL \cite{PGL} of the Selberg averaging:

\begin{align}
\Big< f \Big>^{{\rm BGW}} \ = \ \mathop{\lim\limits_{u,v,N\rightarrow\infty}}_{u+v+2\beta N \equiv \beta n + \beta - 1 } \left( \dfrac{\Big< f \Big>^{{\rm Selb}}}{(uN+\beta N^2)^{\deg f}} \right)
\end{align}
\smallskip\\
or, equivalently, as the unitary integral average (\ref{PGLa}).

Let us describe various averages of the Jack polynomials in these ensembles.

\subsection*{1-Jack average}

\subsubsection*{Selberg model}

The average of single Jack polynomial in the Selberg model has the form

\begin{align}
\Big< J_A(p) \Big>^{{\rm Selb}} \ = \ J_Y\big(\delta_{k,1}\big) \dfrac{[N]_{Y}[u + N\beta + 1 - \beta]_Y}{[u+v+2N\beta + 2 - 2\beta]_Y}
\end{align}
\smallskip\\
where the following notation is used:

\begin{align}
[x]_Y = \prod\limits_{(i,j) \in Y} \Big( x - \beta ( i - 1) + (j - 1) \Big)
\end{align}
\smallskip\\
This Kadell formula is proved in \cite{Kadell}.

\subsubsection*{BGW model}

The average of single Jack polynomial in the BGW model has the form

\begin{align}
\Big< J_A(p) \Big>^{{\rm BGW}} \ = \ J_Y\big(\delta_{k,1}\big) \dfrac{1}{[\beta n + 1 - \beta]_Y}
\end{align}
\smallskip\\
This formula directly follows from the PGL of the Kadell formula.

\pagebreak

\subsection*{2-Jack average}

\subsubsection*{Selberg model}

The average of product of two Jack polynomials in the Selberg model is known in the form

\begin{align*}
\Big< J_A(p + w) J_B(p) \Big>^{{\rm Selb}} \ = \ \dfrac{1}{{\rm Norm_{\beta}(u,v,N)}} \dfrac{[v+N\beta+1-\beta]_A [u+N\beta+1-\beta]_B}{[N \beta]_A
[u+v+N\beta+2-2\beta]_B} \times \emph{}
\end{align*}
\begin{align}
\emph{} \times
\dfrac{\prod\limits_{i<j}^{N} \Big( A_i - A_j + (j-i)\beta\Big)_{\beta} \prod\limits_{i<j}^{N} \Big( B_i - B_j + (j-i)\beta\Big)_{\beta} }
{\prod\limits_{i,j}^{N} \Big( \ u+v+2\beta N + 2+A_i+B_j - (1+i+j)\beta \ \Big)_{\beta}}
\label{Kadell2pt}
\end{align}
\smallskip\\
with the $A,B$-independent normalization constant ${\rm Norm}_{\beta}(u,v,N)$ determined from
$\Big< 1 \Big> = 1$, and
with the conventional Pochhammer symbol $(x)_{\beta}$ defined as
\begin{align}
(x)_{\beta} = \dfrac{\Gamma(x+\beta)}{\Gamma(x)} = x(x+1)\ldots(x+\beta-1)
\end{align}
\smallskip\\
As usual, $A_i$ in (\ref{Kadell2pt}) denotes the height of $i$-th coloumn in the diagram $A$.
The shift $w = (v + 1 - \beta)/\beta$ is essential for the correlator to factorize.
This formula is proved (at least for zero shift, $w = 0$) in \cite{Kadell2}.

Note that eq.(\ref{Kadell2pt}) contains only the heights of Young diagrams, $A_i$ and $B_j$,
while the Nekrasov functions contain also the heights of transposed diagrams like $A^{\prime}$.
The transposed diagrams can be obtained by the following identity:
\begin{align}
j^{(\beta)}_{A}(-p/\beta) = (-1)^{|A|} j^{(1/\beta)}_{A^\prime}(p)
\end{align}
\smallskip\\
which for $\beta = 1$ turns into
\begin{align}
\beta = 1: \ \ \ \ \ \chi_{A}(-p) = (-1)^{|A|} \chi_{A^\prime}(p)
\label{Inversion}
\end{align}
\smallskip\\
and is used below in eq.(\ref{AppendixEq1}).

\subsubsection*{BGW model}

In the PGL only the denominator in the last factor survives in (\ref{Kadell2pt}).
The result can be written as

$$
\big< j_A(p)j_B(p) \big>^{{\rm BGW}}\ =
\prod_{i = 1}^{L_A}
\frac{\Gamma\Big(z +1-(i+L_B)\beta\Big)}{\Gamma\Big(z+1+A_i-(i+L_B)\beta\Big)}
\cdot \prod_{j = 1}^{L_B}
\frac{\Gamma\Big(z+1-(j+L_A)\beta\Big)}{\Gamma\Big(z+1+B_j-(j+L_A)\beta\Big)}
\cdot
$$
\begin{align}
\cdot\prod_{i = 1}^{L_A}\prod_{j = 1}^{L_B}
\frac{\Gamma\Big(z +1 - (i+j-1)\beta\Big)}
{\Gamma\Big(z +1+A_i+B_j-(i+j-1)\beta\Big)}
\cdot\frac{\Gamma\Big(z +1+A_i+B_j-(i+j)\beta\Big)}
{\Gamma\Big(z +1-(i+j)\beta\Big)}
\label{KadellBGW2pt}
\end{align}
where $z = \beta n$, and $L_A, L_B$ are the maximal row lengths of the diagrams $A$ and $B$:
$A = \big(A_1 \geq \ldots \geq A_{L_A}\big)$, $B = \big(B_1 \geq \ldots \geq B_{L_B}\big)$.

\subsubsection*{Case of $\beta = 1$}

Formulas (\ref{Kadell2pt}) and (\ref{KadellBGW2pt}) can look similar to the Nekrasov ones,
however, there is also an important difference: the heights of diagrams in denominator enter not
in combinations like $A_i - j$, which would correspond to Arm- and Leg-lengths, but rather in
combinations like $A_i - \beta j$: very much different from the Nekrasov side. Clearly, this
difference disappears when $\beta = 1$, and above eq.(\ref{Kadell2pt}) can be reduced to

\begin{align}
\boxed{
\Big< \chi_{A}\big( - v - p_k \big) \ \chi_B\big( p_k \big) \Big>^{{\rm Selb}} = \dfrac{(-1)^{|A|+|B|} [-v-N]_A [-u-v-N]_{A} [u+N]_A [N]_B}{G_{AA}(0)G_{AB}(-2N-u-v)G_{BA}(2N+u+v)G_{BB}(0)}
}
\label{AppendixEq1}
\end{align}
\smallskip\\
where the both sides are taken at $\beta = 1$, and the function $G$ has the form

\begin{align}
G_{AB}(x) = \prod\limits_{(i,j) \in A} \Big( x + \beta {\rm Arm}_A(i,j) + {\rm Leg}_B(i,j) + 1 \Big)
\end{align}
\smallskip\\
Recalling that the gauge contribution to the Nekrasov functions has the form

\begin{align}
g_{AB}(x) = G_{AB}(x) G_{AB}(x + \beta - 1)
\label{gGG}
\end{align}
\smallskip\\
one can see that eq.(\ref{AppendixEq1}) then directly implies that

\begin{align}
N_{A,B} = \Big< \chi_{A}\big( - p_k - v_+ \big) \ \chi_B\big(p_k\big) \Big>_+ \ \ \Big< \chi_A\big( p_k \big) \ \chi_{B}\big( -  p_k - v_- \big) \Big>_-
\label{Nek4char}
\end{align}
\smallskip\\
which is the main identity we use in section 3. Note that we wrote eq.(\ref{AppendixEq1}),
and its PGL eq. (\ref{AppendixEq2}) below, in terms of $\chi_A(-p)$ instead of $\chi_A(p)$,
because this is what we need to establish the relation with the DF integral.
Eqs.(\ref{Kadell2pt}) and (\ref{KadellBGW2pt}) involve $\chi_A(p)$, but because of
eq.(\ref{Inversion}), there is essentially no difference at $\beta = 1$.

\subsubsection*{The PGL of the $\beta = 1$ case}

From above eq.(\ref{AppendixEq1}) it follows that

\begin{align}
\boxed{
\Big< \chi_{A}\big( - p_k \big) \ \chi_B\big( p_k \big) \Big>^{{\rm BGW}} = \dfrac{(-1)^{|A|+|B|} }{G_{AA}(0)G_{AB}(-n)G_{BA}(n)G_{BB}(0)}
}
\label{AppendixEq2}
\end{align}
\smallskip\\
where the both sides are taken at $\beta = 1$. Eq.(\ref{gGG}) then implies a BGW counterpart of
(\ref{Nek4char}):

\begin{align}
\Big< \chi_{A}\big( - \tr U^k \big) \ \chi_B\big(\tr U^k\big) \Big>^{BGW}_+ \ \ \Big< \chi_A\big( \tr {\widetilde U}^k \big) \ \chi_{B}\big( -  \tr {\widetilde U}^k \big) \Big>^{BGW}_- = N^{{\rm pure}}_{A,B} \Big|_{\epsilon_1 + \epsilon_2 = 0}
\end{align}
\smallskip\\
which is the main identity we use in section 2.

\pagebreak

\subsection*{BGW multiplication on Young diagrams}

\begin{wrapfigure}{l}{160pt}
  \begin{center}
\vspace{-5ex} \includegraphics[width=150pt]{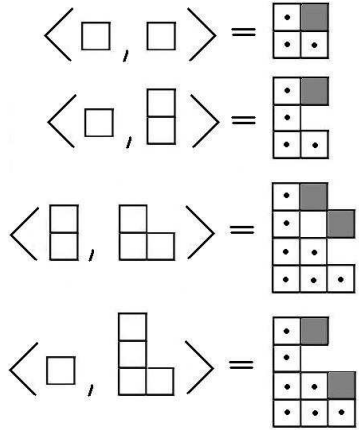}
  \end{center}
\vspace{-5ex}
  \caption{Pictorial representation of the BGW averages: poles are shown as white cells with dots,
  zeroes as grey cells.}
\end{wrapfigure}Clearly all the poles and zeroes of (\ref{KadellBGW2pt}) belong to the first
quadrant (where also zeroes of the Kac determinant are located) and lie in a rectangular
with the length $L_A+L_B$ and the height $H_A+H_B$, where $H_A = A_1$
is the maximal height of the Young diagram $A$.

\begin{figure}
  \begin{center}
\vspace{-5ex} \includegraphics[width=150pt]{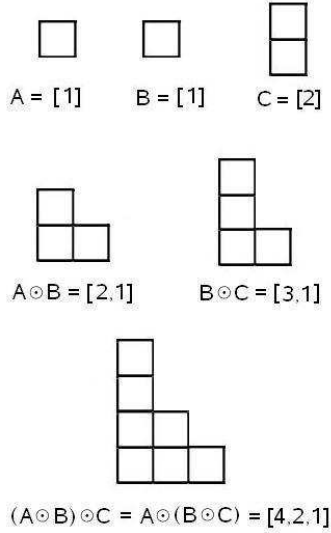}
  \end{center}
  \caption{Multiplication on Young diagrams inspired by study of the BGW correlators.
  The law is simple: for any pair of Young diagrams $A,B$ their "product" $A \odot B$ is equal
  to the Young diagram formed by positions of poles of the correlator $<J_A J_B>$ in the BGW
  model. All known examples suggest that this multiplication is associative.}
\end{figure}

A puzzling observation about these poles is that they always form a new Young diagram, in
particular, there are no multiplicities. Zeroes lie over this newly emerging diagram
(denoted $A \odot B$ in what follows) but form a rather strange configuration.
This issue will be discussed elsewhere, here we just
illustrate it with the two particular examples:

\begin{align}
\hspace{15ex} \big< j_1(p)j_{1}(p) \big>^{{\rm BGW}}\ \sim \dfrac{(2,2)}{(1,1)(1,2)(2,1)}
\end{align}

\begin{align}
\hspace{10ex} \big< j_2(p)j_{21}(p) \big>^{{\rm BGW}}\ \sim \dfrac{(4,2)(3,3)}{(1,1)(2,1)(3,1)(4,1)(1,2)(2,2)(1,3)}
\vspace{10ex}
\end{align}
where $z$-independent normalization prefactors are omitted. Poles (white cells with dots) and
zeroes (grey cells) of these correlators are shown in Figure 13. Note that the poles form a new
Young diagram, a property not shared by the Nekrasov functions, and the same happens for all
choices of $A,B$. Thus, the BGW averaging allows one to define a new amusing commutative
"multiplication" $A,B \mapsto A \odot B$ on Young diagrams. Moreover, this multiplication
seems to be associative! It does not, however, preserve the size-grading:
$|A \odot B| \geq |A| + |B|$.

\pagebreak

\subsection*{Alternative 2-Jack average}

\subsection*{Selberg model}

A somewhat similar, still different 2-Jack correlator was recently considered in \cite{AlbaLit}.
The main difference is the inverse powers $p_{-k} = \sum z_i^{-k}$ in the second argument:

\begin{align}
\Big< J_{A}\big( v + p_k \big) \ J_B\big( p_{-k} \big) \Big>^{{\rm Selb}} \sim G_{AB}(u+v+N\beta+1-\beta) G_{BA}(-u-v-N\beta+2\beta-2)
\label{AlbaEq}
\end{align}
\smallskip\\
Note that $G_{AB}$ functions appear here not in the denominator, but in the numerator, thus,
the r.h.s. of (\ref{AlbaEq}) has no poles. The proportionality coefficient in (\ref{AlbaEq})
depends only on $A$ and $B$ independently: it is equal to

\begin{align}
J_{A}\big( \delta_{k,1} \big) \left( -\dfrac{1}{\beta} \right)^{|A|} \dfrac{[v + N\beta + 1-\beta]_A }{[u+v+2N\beta+2-2\beta]_A} \cdot J_{B}\big( \delta_{k,1} \big)  \left( -\dfrac{1}{\beta} \right)^{|B|}  \dfrac{[N\beta]_B}{[-u]_B}
\end{align}
\smallskip\\
In result, eq.(\ref{AlbaEq}) is of less direct use for the purpose of AGT proof than our
eq.(\ref{AppendixEq1}), however, ref.\cite{AlbaLit} suggests a more involved project with the use of
this formula.

\subsection*{BGW model}

The r.h.s. of (\ref{AlbaEq}) depends {\it not} on the BGW variable
$z = \beta n = u + v + 2\beta N + 1 - \beta$, which is the only combination left finite
in this limit. Thus, the r.h.s. of (\ref{AlbaEq}) becomes trivial in the PGL.

\newpage

\pagestyle{empty}

\begin{sidewaystable}
\centering
\begin{tabular}{|c|c|c|c|cc}
\hline
$A$ & $B$ & $S_{AB} = \Big< j_{A}\big( p_k \big) \ j_B\big( p_k \big) \Big>^{{\rm BGW}}_{\pm}$ & The set of factors &  $\mbox{At } (\epsilon_1,\epsilon_2) = (\hbar,-\hbar)$ & \rule{0pt}{9mm} \\
& & & \\
\hline
$ [1] $&$ [] $&$
\dfrac{g}{z-\epsilon_{2}-\epsilon_{1}}
$&$
g \cdot \dfrac{\varnothing}{(1,1)}
$&$
\dfrac{\hbar}{z}
$&$
\rule{0pt}{9mm} $\\
$ [] $&$ [1] $&$
\dfrac{g}{z-\epsilon_{2}-\epsilon_{1}}
$&$
g \cdot \dfrac{\varnothing}{(1,1)}
$&$
\dfrac{\hbar}{z}
$&$
\rule{0pt}{9mm} $\\
$ [2] $&$ [] $&$
\sqrt{\dfrac{\epsilon_1}{2 \epsilon_{12}}} \cdot \dfrac{-g^2}{(z-\epsilon_{2}-\epsilon_{1}) (z-\epsilon_{1}-2 \epsilon_{2})}
$&$
-g^2 \sqrt{\dfrac{\epsilon_1}{2 \epsilon_{12}}} \cdot \dfrac{\varnothing}{(1,1)(1,2)}
$&$
\dfrac{\hbar^2}{2z(z+\hbar)}
$&$
\rule{0pt}{9mm} $\\
$ [1,1] $&$ [] $&$
\sqrt{\dfrac{-\epsilon_2}{2 \epsilon_{12}}} \cdot \dfrac{-g^2}{(z-\epsilon_{2}-\epsilon_{1}) (z-2 \epsilon_{1}-\epsilon_{2})}
$&$
-g^2 \sqrt{\dfrac{-\epsilon_2}{2 \epsilon_{12}}} \cdot \dfrac{\varnothing}{(1,1)(2,1)}
$&$
\dfrac{\hbar^2}{2z(z-\hbar)}
$&$
\rule{0pt}{9mm} $\\
$ [1] $&$ [1] $&$
\dfrac{-g^2(z-2 \epsilon_{1}-2 \epsilon_{2})}{(z-2 \epsilon_{1}-\epsilon_{2}) (z-\epsilon_{1}-2 \epsilon_{2}) (z-\epsilon_{2}-\epsilon_{1})}
$&$
-g^2 \cdot \dfrac{(2,2)}{(1,1)(2,1)(1,2)}
$&$
\dfrac{\hbar^2}{(z-\hbar)(z+\hbar)}
$&$
\rule{0pt}{9mm} $\\
$ [] $&$ [2] $&$
\sqrt{\dfrac{\epsilon_1}{2 \epsilon_{12}}} \cdot \dfrac{-g^2}{(z-\epsilon_{2}-\epsilon_{1}) (z-\epsilon_{1}-2 \epsilon_{2})}
$&$
-g^2 \sqrt{\dfrac{\epsilon_1}{2 \epsilon_{12}}} \cdot \dfrac{\varnothing}{(1,1)(1,2)}
$&$
\dfrac{\hbar^2}{2z(z+\hbar)}
$&$
\rule{0pt}{9mm} $\\
$ [] $&$ [1,1] $&$
\sqrt{\dfrac{-\epsilon_2}{2 \epsilon_{12}}} \cdot \dfrac{-g^2}{(z-\epsilon_{2}-\epsilon_{1}) (z-2 \epsilon_{1}-\epsilon_{2})}
$&$
-g^2 \sqrt{\dfrac{-\epsilon_2}{2 \epsilon_{12}}} \cdot \dfrac{\varnothing}{(1,1)(2,1)}
$&$
\dfrac{\hbar^2}{2z(z-\hbar)}
$&$
\rule{0pt}{9mm} $\\
\end{tabular}
\vspace{2ex}
\caption[]{Correlators $S_{AB}$ in the "+" and "-" BGW models, at levels $1$ and $2$. Here $z \equiv \pm 2 a$, $g = \sqrt{-\epsilon_1\epsilon_2}$ and $\epsilon_{12}=\epsilon_1-\epsilon_2$. }
\end{sidewaystable}

\begin{sidewaystable}
\centering
\begin{tabular}{|c|c|c|c|cc}
\hline
$A$ & $B$ & $R_{AB} = \Big< j_{A}\big( -p_k \big) \ j_B\big( p_k \big) \Big>^{{\rm BGW}}_{\pm}$ & The set of factors &  $\mbox{At } (\epsilon_1,\epsilon_2) = (\hbar,-\hbar)$ & \rule{0pt}{9mm} \\
& & & \\
\hline
$ [1] $&$ [] $&$
\dfrac{-g}{z-\epsilon_{1}-\epsilon_{2}}
$&$
-g \cdot \dfrac{\varnothing}{(1,1)}
$&$
\dfrac{-\hbar}{z}
$&$
\rule{0pt}{9mm} $\\
$ [] $&$ [1] $&$
\dfrac{g}{z-\epsilon_{1}-\epsilon_{2}}
$&$
g \cdot \dfrac{\varnothing}{(1,1)}
$&$
\dfrac{\hbar}{z}
$&$
\rule{0pt}{9mm} $\\
$ [2] $&$ [] $&$
\sqrt{\dfrac{\epsilon_1}{2 \epsilon_{12}}} \cdot \dfrac{-g^2 (z-2 \epsilon_{1}-3 \epsilon_{2})}{(z-\epsilon_{1}-2 \epsilon_{2})(z-2 \epsilon_{1}-\epsilon_{2}) (z-\epsilon_{1}-\epsilon_{2})}
$&$
-g^2 \sqrt{\dfrac{\epsilon_1}{2 \epsilon_{12}}} \cdot \dfrac{(2,3)}{(1,1)(2,1)(1,2)}
$&$
\dfrac{\hbar^2}{2z(z-\hbar)}
$&$
\rule{0pt}{9mm} $\\
$ [1,1] $&$ [] $&$
\sqrt{\dfrac{-\epsilon_2}{2 \epsilon_{12}}} \cdot \dfrac{-g^2 (z-3 \epsilon_{1}-2 \epsilon_{2})}{(z-\epsilon_{1}-2 \epsilon_{2}) (z-2 \epsilon_{1}-\epsilon_{2}) (z-\epsilon_{1}-\epsilon_{2})}
$&$
-g^2 \sqrt{\dfrac{-\epsilon_2}{2 \epsilon_{12}}} \cdot \dfrac{(3,2)}{(1,1)(2,1)(1,2)}
$&$
\dfrac{\hbar^2}{2z(z+\hbar)}
$&$
\rule{0pt}{9mm} $\\
$ [1] $&$ [1] $&$
\dfrac{g^2 (z-2 \epsilon_{1}-2 \epsilon_{2})}{(z-\epsilon_{1}-2 \epsilon_{2}) (z-2 \epsilon_{1}-\epsilon_{2}) (z-\epsilon_{1}-\epsilon_{2})}
$&$
g^2 \cdot \dfrac{(2,2)}{(1,1)(2,1)(1,2)}
$&$
\dfrac{-\hbar^2}{(z-\hbar)(z+\hbar)}
$&$
\rule{0pt}{9mm} $\\
$ [] $&$ [2] $&$
\sqrt{\dfrac{\epsilon_1}{2 \epsilon_{12}}} \cdot \dfrac{-g^2}{(z-\epsilon_{1}-\epsilon_{2}) (z-\epsilon_{1}-2 \epsilon_{2})}
$&$
-g^2 \sqrt{\dfrac{\epsilon_1}{2 \epsilon_{12}}} \cdot \dfrac{\varnothing}{(1,1)(1,2)}
$&$
\dfrac{\hbar^2}{2z(z+\hbar)}
$&$
\rule{0pt}{9mm} $\\
$ [] $&$ [1,1] $&$
\sqrt{\dfrac{-\epsilon_2}{2 \epsilon_{12}}} \cdot \dfrac{-g^2}{(z-\epsilon_{1}-\epsilon_{2}) (z-2 \epsilon_{1}-\epsilon_{2})}
$&$
-g^2 \sqrt{\dfrac{-\epsilon_2}{2 \epsilon_{12}}} \cdot \dfrac{\varnothing}{(1,1)(2,1)}
$&$
\dfrac{\hbar^2}{2z(z-\hbar)}
$&$
\rule{0pt}{9mm} $\\
\end{tabular}
\vspace{2ex}
\caption[]{Correlators $R_{AB}$ in the "+" and "-" BGW models, at levels $1$ and $2$. Here $z \equiv \pm 2 a$, $g = \sqrt{-\epsilon_1\epsilon_2}$ and $\epsilon_{12}=\epsilon_1-\epsilon_2$. }
\end{sidewaystable}

\begin{sidewaystable}
\centering
\begin{tabular}{|c|c|c|c|cc}
\hline
$A$ & $B$ & $Q_{AB} = \Big< j_{A}\big( -p_k \big) \ j_B\big( - p_k \big) \Big>^{{\rm BGW}}_{\pm}$ & The set of factors &  $\mbox{At } (\epsilon_1,\epsilon_2) = (\hbar,-\hbar)$ & \rule{0pt}{9mm} \\
& & & \\
\hline
$ [1] $&$ [] $&$
\dfrac{-g}{z-\epsilon_{1}-\epsilon_{2}}
$&$
-g \cdot \dfrac{\varnothing}{(1,1)}
$&$
\dfrac{-\hbar}{z}
$&$
\rule{0pt}{9mm} $\\
$ [] $&$ [1] $&$
\dfrac{-g}{z-\epsilon_{1}-\epsilon_{2}}
$&$
-g \cdot \dfrac{\varnothing}{(1,1)}
$&$
\dfrac{-\hbar}{z}
$&$
\rule{0pt}{9mm} $\\
$ [2] $&$ [] $&$
\sqrt{\dfrac{\epsilon_1}{2 \epsilon_{12}}} \cdot \dfrac{-g^2(z-2 \epsilon_{1}-3 \epsilon_{2})}{(z-\epsilon_{1}-2 \epsilon_{2})(z-2 \epsilon_{1}-\epsilon_{2}) (z-\epsilon_{1}-\epsilon_{2})}
$&$
-g^2 \sqrt{\dfrac{\epsilon_1}{2 \epsilon_{12}}} \cdot \dfrac{(2,3)}{(1,1)(2,1)(1,2)}
$&$
\dfrac{\hbar^2}{2z(z-\hbar)}
$&$
\rule{0pt}{9mm} $\\
$ [1,1] $&$ [] $&$
\sqrt{\dfrac{-\epsilon_2}{2 \epsilon_{12}}} \cdot \dfrac{-g^2(z-3 \epsilon_{1}-2 \epsilon_{2})}{(z-\epsilon_{1}-2 \epsilon_{2}) (z-2 \epsilon_{1}-\epsilon_{2}) (z-\epsilon_{1}-\epsilon_{2})}
$&$
-g^2 \sqrt{\dfrac{-\epsilon_2}{2 \epsilon_{12}}} \cdot \dfrac{(3,2)}{(1,1)(2,1)(1,2)}
$&$
\dfrac{\hbar^2}{2z(z+\hbar)}
$&$
\rule{0pt}{9mm} $\\
$ [1] $&$ [1] $&$
\dfrac{-g^2(z-2 \epsilon_{1}-2 \epsilon_{2})}{(z-\epsilon_{1}-2 \epsilon_{2}) (z-2 \epsilon_{1}-\epsilon_{2}) (z-\epsilon_{1}-\epsilon_{2})}
$&$
-g^2 \sqrt{\dfrac{\epsilon_1}{2 \epsilon_{12}}} \cdot \dfrac{(2,2)}{(1,1)(2,1)(1,2)}
$&$
\dfrac{\hbar^2}{(z-\hbar)(z+\hbar)}
$&$
\rule{0pt}{9mm} $\\
$ [] $&$ [2] $&$
\sqrt{\dfrac{\epsilon_1}{2 \epsilon_{12}}} \cdot \dfrac{-g^2(z-2 \epsilon_{1}-3 \epsilon_{2})}{(z-\epsilon_{1}-2 \epsilon_{2})(z-2 \epsilon_{1}-\epsilon_{2}) (z-\epsilon_{1}-\epsilon_{2})}
$&$
-g^2 \sqrt{\dfrac{\epsilon_1}{2 \epsilon_{12}}} \cdot \dfrac{(2,3)}{(1,1)(2,1)(1,2)}
$&$
\dfrac{\hbar^2}{2z(z-\hbar)}
$&$
\rule{0pt}{9mm} $\\
$ [] $&$ [1,1] $&$
\sqrt{\dfrac{-\epsilon_2}{2 \epsilon_{12}}} \cdot \dfrac{-g^2(-3 \epsilon_{1}-2 \epsilon_{2}+z)}{2 (z-\epsilon_{1}-2 \epsilon_{2}) (z-2 \epsilon_{1}-\epsilon_{2}) (z-\epsilon_{1}-\epsilon_{2})}
$&$
-g^2 \sqrt{\dfrac{-\epsilon_2}{2 \epsilon_{12}}} \cdot \dfrac{(3,2)}{(1,1)(2,1)(1,2)}
$&$
\dfrac{\hbar^2}{2z(z+\hbar)}
$&$
\rule{0pt}{9mm} $\\
\end{tabular}
\vspace{2ex}
\caption[]{Correlators $Q_{AB}$ in the "+" and "-" BGW models, at levels $1$ and $2$.
Here $z \equiv \pm 2 a$,
$g = \sqrt{-\epsilon_1\epsilon_2}$ and $\epsilon_{12}=\epsilon_1-\epsilon_2$. }
\end{sidewaystable}

\begin{sidewaystable}
\centering
\begin{tabular}{|c|c|cccc}
\hline
$A$ & $B$ & $\Big< j_{A}\big( - w - p_k \big) \ j_B\big( -p_k \big) \Big>^{{\rm Selb}}_{\pm}$ \rule{0pt}{9mm} \\
& & & \\
\hline
$ [1] $&$ [] $&$
\dfrac{1}{\sqrt{\beta}} \dfrac{(v+2-2 \beta+u+N \beta) (N \beta-\beta+1+v)}{(u+v+2 N \beta+2-2 \beta)}
$&$
\rule{0pt}{9mm} $\\
$ [] $&$ [1] $&$
\dfrac{1}{\sqrt{\beta}} \dfrac{(u+N \beta-\beta+1)N \beta}{u+v+2 N \beta+2-2 \beta}
$&$
\rule{0pt}{9mm} $\\
$ [2] $&$ [] $&$
\dfrac{1}{\sqrt{2\beta(\beta+1)}} \dfrac{(v+2-\beta+N \beta) (v+1-\beta+N \beta) (v+3+u-2 \beta+N \beta) (v+2-2 \beta+N \beta+u)}{(u+v+2 N \beta+2-2 \beta) (u+v+2 N \beta+3-2 \beta)}
$&$
\rule{0pt}{9mm} $\\
$ [1,1] $&$ [] $&$
\dfrac{1}{\sqrt{2\beta^2(\beta+1)}} \dfrac{(v+1-2 \beta+N \beta) (v+1+N \beta-\beta) (v+2+u-3 \beta+N \beta) (v+2+u-2 \beta+N \beta)}{(u+v+2 N \beta+2-2 \beta) (u+v+2 N \beta+2-3 \beta)}
$&$
\rule{0pt}{9mm} $\\
$ [1] $&$ [1] $&$
\dfrac{1}{\beta} \dfrac{(v+1+N \beta-\beta) (v+2+u-2 \beta+N \beta) (v+3-3 \beta+2 N \beta+u)(u+N \beta-\beta+1) N\beta }{(u+v+2 N \beta+3-2 \beta) (u+v+2 N \beta+2-2 \beta) (u+v+2 N \beta+2-3 \beta)}
$&$
\rule{0pt}{9mm} $\\
$ [] $&$ [2] $&$
\dfrac{1}{\sqrt{2\beta(\beta+1)}} \dfrac{N\beta (N \beta+1) (u+N \beta-\beta+1) (u+N \beta-\beta+2)}{(u+v+2 N \beta+2-2 \beta) (u+v+2 N \beta+3-2 \beta)}
$&$
\rule{0pt}{9mm} $\\
$ [] $&$ [1,1] $&$
\dfrac{1}{\sqrt{2\beta^2(\beta+1)}} \dfrac{N\beta (N\beta-1) (u+N \beta-\beta+1) (u+N \beta-2 \beta+1)}{(u+v+2 N \beta+2-2 \beta) (u+v+2 N \beta+2-3 \beta)}
$&$
\rule{0pt}{9mm} $\\
\end{tabular}
\vspace{2ex}
\caption[]{The table of correlators in the "+" and "-" Selberg models, at levels $1$ and $2$.
The correlators are written in matrix model notations, where $u,v$ are the parameters of the Selberg
potential, $N$ is the number of eigenvalues, and $\beta = -\epsilon_1/\epsilon_2$ is the
Van-der-Monde power. The shift $w$, which is essential for the correlators to be completely
factorizable, are equal to $w = (v + 1 - \beta)/\beta$. }
\end{sidewaystable}

\begin{sidewaystable}
\centering
\begin{tabular}{|c|c|cccc}
\hline
$A$ & $B$ & $\Big< j_{A}\big( w + p_k \big) \ j_B\big( p_{-k} \big) \Big>^{{\rm Selb}}_{\pm}$ \rule{0pt}{9mm} \\
& & & \\
\hline
$ [1] $&$ [] $&$
\dfrac{1}{\sqrt{\beta}} \dfrac{(1+u+v+N \beta-\beta)N \beta}{u}
$&$
\rule{0pt}{9mm} $\\
$ [] $&$ [1] $&$
\dfrac{1}{\sqrt{\beta}} \dfrac{(v+1-\beta+N \beta) (v+u+N \beta+2-2 \beta)}{\beta (u+v+2 N \beta+2-2 \beta)}
$&$
\rule{0pt}{9mm} $\\
$ [2] $&$ [] $&$
\dfrac{1}{\sqrt{2\beta(\beta+1)}} \dfrac{N\beta (N \beta+1) (1+u+v+N \beta-\beta) (u+v+N \beta-\beta)}{u (-1+u)}
$&$
\rule{0pt}{9mm} $\\
$ [1,1] $&$ [] $&$
\dfrac{1}{\sqrt{2\beta^2(\beta+1)}} \dfrac{N\beta (N\beta-\beta) (1+u+v+N \beta) (1+u+v+N \beta-\beta)}{2u (u+\beta)}
$&$
\rule{0pt}{9mm} $\\
$ [1] $&$ [1] $&$
\dfrac{1}{\beta} \dfrac{(v+1-\beta+N \beta) (u+v+1+N \beta-2 \beta) (2+u+v+N \beta-\beta) N\beta}{\beta u (u+v+2 N \beta+2-2 \beta)}
$&$
\rule{0pt}{9mm} $\\
$ [] $&$ [2] $&$
\dfrac{1}{\sqrt{2\beta(\beta+1)}} \dfrac{(v+1-\beta+N \beta) (2+v+N \beta-\beta) (v+u+N \beta+2-2 \beta) (u+v+3+N \beta-2 \beta)}{(u+v+2 N \beta+2-2 \beta) (3+u+v+2 N \beta-2 \beta)}
$&$
\rule{0pt}{9mm} $\\
$ [] $&$ [1,1] $&$
\dfrac{1}{\sqrt{2\beta^2(\beta+1)}} \dfrac{(v+1-\beta+N \beta) (1+v+N \beta-2 \beta) (u+v+2+N \beta-3 \beta) (v+u+N \beta+2-2 \beta)}{(u+v+2 N \beta+2-2 \beta) (2+u+v+2 N \beta-3 \beta)}
$&$
\rule{0pt}{9mm} $\\
\end{tabular}
\vspace{2ex}
\caption[]{The Selberg correlators considered in [...] -- with inverse $p_{-k}$
in the second Jack polynomial -- at levels $|A| + |B| = 1$ and $|A| + |B| = 2$.
Again, $u,v$ are the parameters of the
Selberg potential, $N$ is the number of eigenvalues, and $\beta = -\epsilon_1/\epsilon_2$
is the Van-der-Monde power. The shift $w$, which is essential for the correlators to be
completely factorizable, equals $w = (v + 1 - \beta)/\beta$. }
\end{sidewaystable}

\begin{sidewaystable}
\centering
\begin{tabular}{|c|c|c|ccc}
\hline
$A$ & $B$ & $S_{AB} = \Big< j_{A}\big( p_k \big) \ j_B\big( p_k \big) \Big>^{{\rm BGW}}_{\pm}$ & $\mbox{At } (\epsilon_1,\epsilon_2) = (\hbar,-\hbar)$ & \rule{0pt}{9mm} \\
& & & \\
\hline
$ [3] $&$ [] $&$
\sqrt{\dfrac{\epsilon_1^2}{6\epsilon_{12}\epsilon_{122}}} \cdot \dfrac{-g^3}{(z-\epsilon_{2}-\epsilon_{1}) (z-\epsilon_{1}-2 \epsilon_{2}) (z-\epsilon_{1}-3 \epsilon_{2})}
$&$
\dfrac{\hbar^3}{6z(z+\hbar)(z+2\hbar)}
$&$
\rule{0pt}{9mm} $\\
$ [2,1] $&$ [] $&$
\sqrt{\dfrac{-\epsilon_1\epsilon_2}{\epsilon_{112}\epsilon_{122}}} \cdot \dfrac{-g^3}{(z-\epsilon_{2}-\epsilon_{1}) (z-\epsilon_{1}-2 \epsilon_{2}) (z-2 \epsilon_{1}-\epsilon_{2})}
$&$
\dfrac{\hbar^3}{3z(z-\hbar)(z+\hbar)}
$&$
\rule{0pt}{9mm} $\\
$ [1,1,1] $&$ [] $&$
\sqrt{\dfrac{\epsilon_2^2}{6\epsilon_{12}\epsilon_{112}}} \cdot \dfrac{-g^3}{(z-\epsilon_{2}-\epsilon_{1}) (z-2 \epsilon_{1}-\epsilon_{2}) (z-3 \epsilon_{1}-\epsilon_{2})}
$&$
\dfrac{\hbar^3}{6z(z-\hbar)(z-2\hbar)}
$&$
\rule{0pt}{9mm} $\\
$ [2] $&$ [1] $&$
\sqrt{\dfrac{\epsilon_1}{2\epsilon_{12}}} \cdot \dfrac{-g^3(z-2 \epsilon_{1}-3 \epsilon_{2})}{(z-2 \epsilon_{1}-\epsilon_{2}) (z-\epsilon_{1}-3 \epsilon_{2}) (z-\epsilon_{1}-2 \epsilon_{2}) (z-\epsilon_{2}-\epsilon_{1})}
$&$
\dfrac{\hbar^3}{2z(z-\hbar)(z+2\hbar)}
$&$
\rule{0pt}{9mm} $\\
$ [1,1] $&$ [1] $&$
\sqrt{\dfrac{-\epsilon_2}{2\epsilon_{12}}} \cdot \dfrac{-g^3(z-3 \epsilon_{1}-2 \epsilon_{2})}{(z-\epsilon_{1}-2 \epsilon_{2}) (z-3 \epsilon_{1}-\epsilon_{2}) (z-2 \epsilon_{1}-\epsilon_{2}) (z-\epsilon_{2}-\epsilon_{1})}
$&$
\dfrac{\hbar^3}{2z(z+\hbar)(z-2\hbar)}
$&$
\rule{0pt}{9mm} $\\
$ [1] $&$ [2] $&$
\sqrt{\dfrac{\epsilon_1}{2\epsilon_{12}}} \cdot \dfrac{-g^3(z-2 \epsilon_{1}-3 \epsilon_{2})}{(z-2 \epsilon_{1}-\epsilon_{2}) (z-\epsilon_{1}-3 \epsilon_{2}) (z-\epsilon_{1}-2 \epsilon_{2}) (z-\epsilon_{2}-\epsilon_{1})}
$&$
\dfrac{\hbar^3}{2z(z-\hbar)(z+2\hbar)}
$&$
\rule{0pt}{9mm} $\\
$ [1] $&$ [1,1] $&$
\sqrt{\dfrac{-\epsilon_2}{2\epsilon_{12}}} \cdot \dfrac{-g^3(z-3 \epsilon_{1}-2 \epsilon_{2})}{(z-\epsilon_{1}-2 \epsilon_{2}) (z-3 \epsilon_{1}-\epsilon_{2}) (z-2 \epsilon_{1}-\epsilon_{2}) (z-\epsilon_{2}-\epsilon_{1})}
$&$
\dfrac{\hbar^3}{2z(z+\hbar)(z-2\hbar)}
$&$
\rule{0pt}{9mm} $\\
$ [] $&$ [3] $&$
\sqrt{\dfrac{\epsilon_1^2}{6\epsilon_{12}\epsilon_{122}}} \cdot \dfrac{-g^3}{(z-\epsilon_{2}-\epsilon_{1}) (z-\epsilon_{1}-2 \epsilon_{2}) (z-\epsilon_{1}-3 \epsilon_{2})}
$&$
\dfrac{\hbar^3}{6z(z+\hbar)(z+2\hbar)}
$&$
\rule{0pt}{9mm} $\\
$ [] $&$ [2,1] $&$
\sqrt{\dfrac{-\epsilon_1\epsilon_2}{\epsilon_{112}\epsilon_{122}}} \cdot \dfrac{-g^3}{(z-\epsilon_{2}-\epsilon_{1}) (z-\epsilon_{1}-2 \epsilon_{2}) (z-2 \epsilon_{1}-\epsilon_{2})}
$&$
\dfrac{\hbar^3}{3z(z-\hbar)(z+\hbar)}
$&$
\rule{0pt}{9mm} $\\
$ [] $&$ [1,1,1] $&$
\sqrt{\dfrac{\epsilon_2^2}{6\epsilon_{12}\epsilon_{112}}} \cdot \dfrac{-g^3}{(z-\epsilon_{2}-\epsilon_{1}) (z-2 \epsilon_{1}-\epsilon_{2}) (z-3 \epsilon_{1}-\epsilon_{2})}
$&$
\dfrac{\hbar^3}{6z(z-\hbar)(z-2\hbar)}
$&$
\rule{0pt}{9mm} $\\
\end{tabular}
\vspace{2ex}
\caption[]{Correlators $S_{AB}$ in the "+" and "-" BGW models, level $3$. Here $z \equiv \pm 2 a$,
$g = \sqrt{-\epsilon_1\epsilon_2}, \epsilon_{12}=\epsilon_1-\epsilon_2,
\epsilon_{112} = 2\epsilon_1-\epsilon_2, \epsilon_{122} = \epsilon_1-2\epsilon_2$.}
\end{sidewaystable}

\begin{sidewaystable}
\centering
\begin{tabular}{|c|c|c|ccc}
\hline
$A$ & $B$ & $R_{AB} = \Big< j_{A}\big( - p_k \big) \ j_B\big( p_k \big) \Big>^{{\rm BGW}}_{\pm}$ & $\mbox{At } (\epsilon_1,\epsilon_2) = (\hbar,-\hbar)$ & \rule{0pt}{9mm} \\
& & & \\
\hline
$ [3] $&$ [] $&$
\sqrt{\dfrac{\epsilon_1^2}{6\epsilon_{12}\epsilon_{122}}} \cdot \dfrac{g^3(6 \epsilon_{1}^2+23 \epsilon_{1} \epsilon_{2}-5 \epsilon_{1} z+19 \epsilon_{2}^2-8 \epsilon_{2} z+z^2)}{(z-\epsilon_{1}-3 \epsilon_{2}) (z-3 \epsilon_{1}-\epsilon_{2}) (z-2 \epsilon_{1}-\epsilon_{2}) (z-\epsilon_{1}-2 \epsilon_{2}) (z-\epsilon_{1}-\epsilon_{2})}
$&$
\dfrac{-\hbar^3}{6z(z-\hbar)(z-2\hbar)}
$&$
\rule{0pt}{9mm} $\\
$ [2,1] $&$ [] $&$
\sqrt{\dfrac{-\epsilon_1\epsilon_2}{\epsilon_{112}\epsilon_{122}}} \cdot \dfrac{g^3(9 \epsilon_{1}^2-6 \epsilon_{1} z+22 \epsilon_{1} \epsilon_{2}+9 \epsilon_{2}^2+z^2-6 \epsilon_{2} z) }{(z-\epsilon_{1}-3 \epsilon_{2}) (z-3 \epsilon_{1}-\epsilon_{2})(z-2 \epsilon_{1}-\epsilon_{2}) (z-\epsilon_{1}-2 \epsilon_{2}) (z-\epsilon_{1}-\epsilon_{2})}
$&$
\dfrac{-\hbar^3}{3z(z-\hbar)(z+\hbar)}
$&$
\rule{0pt}{9mm} $\\
$ [1,1,1] $&$ [] $&$
\sqrt{\dfrac{\epsilon_2^2}{6\epsilon_{12}\epsilon_{112}}} \cdot \dfrac{g^3(19 \epsilon_{1}^2+23 \epsilon_{1} \epsilon_{2}-8 \epsilon_{1} z+z^2-5 \epsilon_{2} z+6 \epsilon_{2}^2)}{(z-\epsilon_{1}-3 \epsilon_{2}) (z-3 \epsilon_{1}-\epsilon_{2}) (z-2 \epsilon_{1}-\epsilon_{2}) (z-\epsilon_{1}-2 \epsilon_{2}) (z-\epsilon_{1}-\epsilon_{2})}
$&$
\dfrac{-\hbar^3}{6z(z+\hbar)(z+2\hbar)}
$&$
\rule{0pt}{9mm} $\\
$ [2] $&$ [1] $&$
\sqrt{\dfrac{\epsilon_1}{2\epsilon_{12}}} \cdot \dfrac{-g^3(6 \epsilon_{1}^2-5 \epsilon_{1} z+17 \epsilon_{1} \epsilon_{2}+9 \epsilon_{2}^2+z^2-6 \epsilon_{2} z) }{(z-\epsilon_{1}-3 \epsilon_{2}) (z-3 \epsilon_{1}-\epsilon_{2})(z-2 \epsilon_{1}-\epsilon_{2}) (z-\epsilon_{1}-2 \epsilon_{2}) (z-\epsilon_{1}-\epsilon_{2})}
$&$
\dfrac{\hbar^3}{2z(z+\hbar)(z-2\hbar)}
$&$
\rule{0pt}{9mm} $\\
$ [1,1] $&$ [1] $&$
\sqrt{\dfrac{-\epsilon_2}{2\epsilon_{12}}} \cdot \dfrac{-g^3(9 \epsilon_{1}^2-6 \epsilon_{1} z+17 \epsilon_{1} \epsilon_{2}+6 \epsilon_{2}^2-5 \epsilon_{2} z+z^2)}{(z-\epsilon_{1}-3 \epsilon_{2}) (z-3 \epsilon_{1}-\epsilon_{2}) (z-2 \epsilon_{1}-\epsilon_{2}) (z-\epsilon_{1}-2 \epsilon_{2}) (z-\epsilon_{1}-\epsilon_{2})}
$&$
\dfrac{\hbar^3}{2z(z-\hbar)(z+2\hbar)}
$&$
\rule{0pt}{9mm} $\\
$ [1] $&$ [2] $&$
\sqrt{\dfrac{\epsilon_1}{2\epsilon_{12}}} \cdot \dfrac{g^3(-2 \epsilon_{1}-3 \epsilon_{2}+z)}{(z-\epsilon_{1}-3 \epsilon_{2}) (z-2 \epsilon_{1}-\epsilon_{2}) (z-\epsilon_{1}-2 \epsilon_{2}) (z-\epsilon_{1}-\epsilon_{2})}
$&$
\dfrac{-\hbar^3}{2z(z-\hbar)(z+2\hbar)}
$&$
\rule{0pt}{9mm} $\\
$ [1] $&$ [1,1] $&$
\sqrt{\dfrac{-\epsilon_2}{2\epsilon_{12}}} \cdot \dfrac{g^3(-3 \epsilon_{1}-2 \epsilon_{2}+z) }{(z-3 \epsilon_{1}-\epsilon_{2}) (z-2 \epsilon_{1}-\epsilon_{2}) (z-\epsilon_{1}-2 \epsilon_{2}) (z-\epsilon_{1}-\epsilon_{2})}
$&$
\dfrac{-\hbar^3}{2z(z+\hbar)(z-2\hbar)}
$&$
\rule{0pt}{9mm} $\\
$ [] $&$ [3] $&$
\sqrt{\dfrac{\epsilon_1^2}{6\epsilon_{12}\epsilon_{122}}} \cdot \dfrac{-g^3}{(z-\epsilon_{1}-\epsilon_{2}) (z-\epsilon_{1}-2 \epsilon_{2}) (z-\epsilon_{1}-3 \epsilon_{2})}
$&$
\dfrac{\hbar^3}{6z(z+\hbar)(z+2\hbar)}
$&$
\rule{0pt}{9mm} $\\
$ [] $&$ [2,1] $&$
\sqrt{\dfrac{-\epsilon_1\epsilon_2}{\epsilon_{112}\epsilon_{122}}} \cdot \dfrac{-g^3}{(z-\epsilon_{1}-\epsilon_{2}) (z-\epsilon_{1}-2 \epsilon_{2}) (z-2 \epsilon_{1}-\epsilon_{2})}
$&$
\dfrac{\hbar^3}{3z(z+\hbar)(z-\hbar)}
$&$
\rule{0pt}{9mm} $\\
$ [] $&$ [1,1,1] $&$
\sqrt{\dfrac{\epsilon_2^2}{6\epsilon_{12}\epsilon_{112}}} \cdot \dfrac{-g^3}{(z-\epsilon_{1}-\epsilon_{2}) (z-2 \epsilon_{1}-\epsilon_{2}) (z-3 \epsilon_{1}-\epsilon_{2})}
$&$
\dfrac{\hbar^3}{6z(z-\hbar)(z-2\hbar)}
$&$
\rule{0pt}{9mm} $\\
\end{tabular}
\vspace{2ex}
\caption[]{Correlators $R_{AB}$ in the "+" and "-" BGW models, level $3$. Here $z \equiv \pm 2 a$,
$g = \sqrt{-\epsilon_1\epsilon_2}, \epsilon_{12}=\epsilon_1-\epsilon_2,
\epsilon_{112} = 2\epsilon_1-\epsilon_2, \epsilon_{122} = \epsilon_1-2\epsilon_2$.}
\end{sidewaystable}

\begin{sidewaystable}
\centering
\begin{tabular}{|c|c|c|ccc}
\hline
$A$ & $B$ & $Q_{AB} = \Big< j_{A}\big( - p_k \big) \ j_B\big( -p_k \big) \Big>^{{\rm BGW}}_{\pm}$ & $\mbox{At } (\epsilon_1,\epsilon_2) = (\hbar,-\hbar)$ & \rule{0pt}{9mm} \\
& & & \\
\hline
$ [3] $&$ [] $&$
\sqrt{\dfrac{\epsilon_1^2}{6\epsilon_{12}\epsilon_{122}}} \cdot \dfrac{g^3(6 \epsilon_{1}^2+23 \epsilon_{1} \epsilon_{2}-5 \epsilon_{1} z-8 \epsilon_{2} z+19 \epsilon_{2}^2+z^2)}{(z-\epsilon_{1}-3 \epsilon_{2}) (-\epsilon_{2}+\epsilon_{1}) (z-3 \epsilon_{1}-\epsilon_{2}) (-2 \epsilon_{2}+\epsilon_{1}) (z-2 \epsilon_{1}-\epsilon_{2}) (z-\epsilon_{1}-2 \epsilon_{2}) (z-\epsilon_{1}-\epsilon_{2})}
$&$
\dfrac{-\hbar^3}{6z(z-\hbar)(z-2\hbar)}
$&$
\rule{0pt}{9mm} $\\
$ [2,1] $&$ [] $&$
\sqrt{\dfrac{-\epsilon_1\epsilon_2}{\epsilon_{112}\epsilon_{122}}} \cdot \dfrac{g^3(9 \epsilon_{1}^2+22 \epsilon_{1} \epsilon_{2}-6 \epsilon_{1} z+9 \epsilon_{2}^2+z^2-6 \epsilon_{2} z)}{(z-\epsilon_{1}-3 \epsilon_{2}) (z-3 \epsilon_{1}-\epsilon_{2}) (-\epsilon_{2}+2 \epsilon_{1}) (z-2 \epsilon_{1}-\epsilon_{2}) (z-\epsilon_{1}-2 \epsilon_{2}) (z-\epsilon_{1}-\epsilon_{2})}
$&$
\dfrac{-\hbar^3}{3z(z+\hbar)(z-\hbar)}
$&$
\rule{0pt}{9mm} $\\
$ [1,1,1] $&$ [] $&$
\sqrt{\dfrac{\epsilon_2^2}{6\epsilon_{12}\epsilon_{112}}} \cdot \dfrac{g^3(19 \epsilon_{1}^2-8 \epsilon_{1} z+23 \epsilon_{1} \epsilon_{2}+z^2+6 \epsilon_{2}^2-5 \epsilon_{2} z)}{(z-\epsilon_{1}-3 \epsilon_{2}) (z-3 \epsilon_{1}-\epsilon_{2}) (z-2 \epsilon_{1}-\epsilon_{2}) (z-\epsilon_{1}-2 \epsilon_{2}) (z-\epsilon_{1}-\epsilon_{2})}
$&$
\dfrac{-\hbar^3}{6z(z+\hbar)(z+2\hbar)}
$&$
\rule{0pt}{9mm} $\\
$ [2] $&$ [1] $&$
\sqrt{\dfrac{\epsilon_1}{2\epsilon_{12}}} \cdot \dfrac{g^3(6 \epsilon_{1}^2+17 \epsilon_{1} \epsilon_{2}-5 \epsilon_{1} z+z^2-6 \epsilon_{2} z+9 \epsilon_{2}^2)}{(z-\epsilon_{1}-3 \epsilon_{2}) (z-3 \epsilon_{1}-\epsilon_{2}) (-\epsilon_{2}+\epsilon_{1}) (z-2 \epsilon_{1}-\epsilon_{2}) (z-\epsilon_{1}-2 \epsilon_{2}) (z-\epsilon_{1}-\epsilon_{2})}
$&$
\dfrac{-\hbar^3}{2z(z+\hbar)(z-2\hbar)}
$&$
\rule{0pt}{9mm} $\\
$ [1,1] $&$ [1] $&$
\sqrt{\dfrac{-\epsilon_2}{2\epsilon_{12}}} \cdot \dfrac{g^3(9 \epsilon_{1}^2+17 \epsilon_{1} \epsilon_{2}-6 \epsilon_{1} z-5 \epsilon_{2} z+z^2+6 \epsilon_{2}^2)}{(z-\epsilon_{1}-3 \epsilon_{2}) (z-3 \epsilon_{1}-\epsilon_{2}) (z-2 \epsilon_{1}-\epsilon_{2}) (z-\epsilon_{1}-2 \epsilon_{2}) (z-\epsilon_{1}-\epsilon_{2})}
$&$
\dfrac{-\hbar^3}{2z(z-\hbar)(z+2\hbar)}
$&$
\rule{0pt}{9mm} $\\
$ [1] $&$ [2] $&$
\sqrt{\dfrac{\epsilon_1}{2\epsilon_{12}}} \cdot \dfrac{g^3(6 \epsilon_{1}^2-5 \epsilon_{1} z+17 \epsilon_{1} \epsilon_{2}+9 \epsilon_{2}^2+z^2-6 \epsilon_{2} z)}{(z-\epsilon_{1}-3 \epsilon_{2}) (z-3 \epsilon_{1}-\epsilon_{2}) (-\epsilon_{2}+\epsilon_{1}) (z-2 \epsilon_{1}-\epsilon_{2}) (z-\epsilon_{1}-2 \epsilon_{2}) (z-\epsilon_{1}-\epsilon_{2})}
$&$
\dfrac{-\hbar^3}{2z(z+\hbar)(z-2\hbar)}
$&$
\rule{0pt}{9mm} $\\
$ [1] $&$ [1,1] $&$
\sqrt{\dfrac{-\epsilon_2}{2\epsilon_{12}}} \cdot \dfrac{g^3(9 \epsilon_{1}^2-6 z \epsilon_{1}+17 \epsilon_{1} \epsilon_{2}+6 \epsilon_{2}^2-5 z \epsilon_{2}+z^2)}{(z-\epsilon_{1}-3 \epsilon_{2}) (z-3 \epsilon_{1}-\epsilon_{2}) (z-2 \epsilon_{1}-\epsilon_{2}) (z-\epsilon_{1}-2 \epsilon_{2}) (z-\epsilon_{1}-\epsilon_{2})}
$&$
\dfrac{-\hbar^3}{2z(z-\hbar)(z+2\hbar)}
$&$
\rule{0pt}{9mm} $\\
$ [] $&$ [3] $&$
\sqrt{\dfrac{\epsilon_1^2}{6\epsilon_{12}\epsilon_{122}}} \cdot \dfrac{g^3(6 \epsilon_{1}^2+23 \epsilon_{1} \epsilon_{2}-5 \epsilon_{1} z+z^2-8 \epsilon_{2} z+19 \epsilon_{2}^2)}{(z-\epsilon_{1}-3 \epsilon_{2}) (-\epsilon_{2}+\epsilon_{1}) (z-3 \epsilon_{1}-\epsilon_{2}) (-2 \epsilon_{2}+\epsilon_{1}) (z-2 \epsilon_{1}-\epsilon_{2}) (z-\epsilon_{1}-2 \epsilon_{2}) (z-\epsilon_{1}-\epsilon_{2})}
$&$
\dfrac{-\hbar^3}{6z(z-\hbar)(z-2\hbar)}
$&$
\rule{0pt}{9mm} $\\
$ [] $&$ [2,1] $&$
\sqrt{\dfrac{-\epsilon_1\epsilon_2}{\epsilon_{112}\epsilon_{122}}} \cdot \dfrac{g^3(9 \epsilon_{1}^2+22 \epsilon_{1} \epsilon_{2}-6 \epsilon_{1} z+9 \epsilon_{2}^2+z^2-6 z \epsilon_{2})}{(z-\epsilon_{1}-3 \epsilon_{2}) (z-3 \epsilon_{1}-\epsilon_{2}) (-\epsilon_{2}+2 \epsilon_{1}) (z-2 \epsilon_{1}-\epsilon_{2}) (z-\epsilon_{1}-2 \epsilon_{2}) (z-\epsilon_{1}-\epsilon_{2})}
$&$
\dfrac{-\hbar^3}{3z(z+\hbar)(z-\hbar)}
$&$
\rule{0pt}{9mm} $\\
$ [] $&$ [1,1,1] $&$
\sqrt{\dfrac{\epsilon_2^2}{6\epsilon_{12}\epsilon_{112}}} \cdot \dfrac{g^3(19 \epsilon_{1}^2+23 \epsilon_{1} \epsilon_{2}-8 z \epsilon_{1}-5 z \epsilon_{2}+z^2+6 \epsilon_{2}^2)}{(z-\epsilon_{1}-3 \epsilon_{2}) (z-3 \epsilon_{1}-\epsilon_{2}) (z-2 \epsilon_{1}-\epsilon_{2}) (z-\epsilon_{1}-2 \epsilon_{2}) (z-\epsilon_{1}-\epsilon_{2})}
$&$
\dfrac{-\hbar^3}{6z(z+\hbar)(z+2\hbar)}
$&$
\rule{0pt}{9mm} $\\
\end{tabular}
\vspace{2ex}
\caption[]{Correlators $Q_{AB}$ in the "+" and "-" BGW models, level $3$. Here $z \equiv \pm 2 a$,
$g = \sqrt{-\epsilon_1\epsilon_2}, \epsilon_{12}=\epsilon_1-\epsilon_2,
\epsilon_{112} = 2\epsilon_1-\epsilon_2, \epsilon_{122} = \epsilon_1-2\epsilon_2$.}
\end{sidewaystable}

\end{document}